\documentclass[12pt]{iopart}

\usepackage[dvips]{graphicx}
\usepackage{dsfont}
\usepackage{subfigure, epsfig}
\usepackage{braket}
\usepackage{hyperref}
\usepackage{bm}
\usepackage{enumerate}
\usepackage{color}
\usepackage{graphicx}

\newcommand{\affiliation}[1]{\address{#1}}
\newcommand{\eqref}[1]{(\ref{#1})}

\newcommand{\tfrac}[2]{\case{#1}{#2}}
\newcommand{\binom}[2]{{#1\choose{#2}}}
\newcommand{\mathbb}[1]{\mathds{#1}}

\begin{document}


\title{Variational-State Quantum Metrology}

%
\author{B\'alint Koczor$^1$, Suguru Endo$^1$, Tyson Jones$^1$, Yuichiro Matsuzaki$^2$, Simon C. Benjamin$^1$}

\affiliation{$^1$Department of Materials, University of Oxford, Parks Road, Oxford OX1 3PH, United Kingdom}

\begin{abstract}
Quantum technologies exploit entanglement to enhance various tasks beyond their
classical limits including computation, communication and measurements.
Quantum metrology aims to increase the precision of a measured quantity
that is estimated in the presence of statistical errors using entangled quantum
states. 
We present a novel approach for finding (near) optimal
states for metrology in the presence of noise, using variational techniques as a tool for efficiently searching
the classically intractable high-dimensional space of quantum states.
We comprehensively explore systems consisting of up to 9 qubits
and find new highly entangled states that are not symmetric under permutations and
non-trivially outperform previously known states up to a constant factor $2$.
We consider a range of environmental noise models; while passive quantum states cannot achieve a fundamentally superior scaling
(as established by prior asymptotic results) we do observe a significant absolute quantum advantage.  
We finally outline a possible experimental setup for variational quantum metrology
 which can be implemented in near-term hardware.
\end{abstract}

\section{Introduction}
Variational quantum algorithms (VQAs) are potentially powerful for
solving various problems using near-term quantum computers
\cite{peruzzo2014variational,kandala2017hardware,moll2018quantum,mcclean2016theory,farhi2014quantum,li2017efficient}. 
These techniques can be implemented on shallow-depth quantum circuits
that depend on external parameters and these parameters are typically optimised
externally by a classical computer.
Moreover, variational quantum  algorithms are expected to be the first applications of
quantum computers that could potentially outperform the best classical computers
in practically relevant tasks.

\begin{figure*}[tb]
	\begin{centering}
		\includegraphics[width=\linewidth]{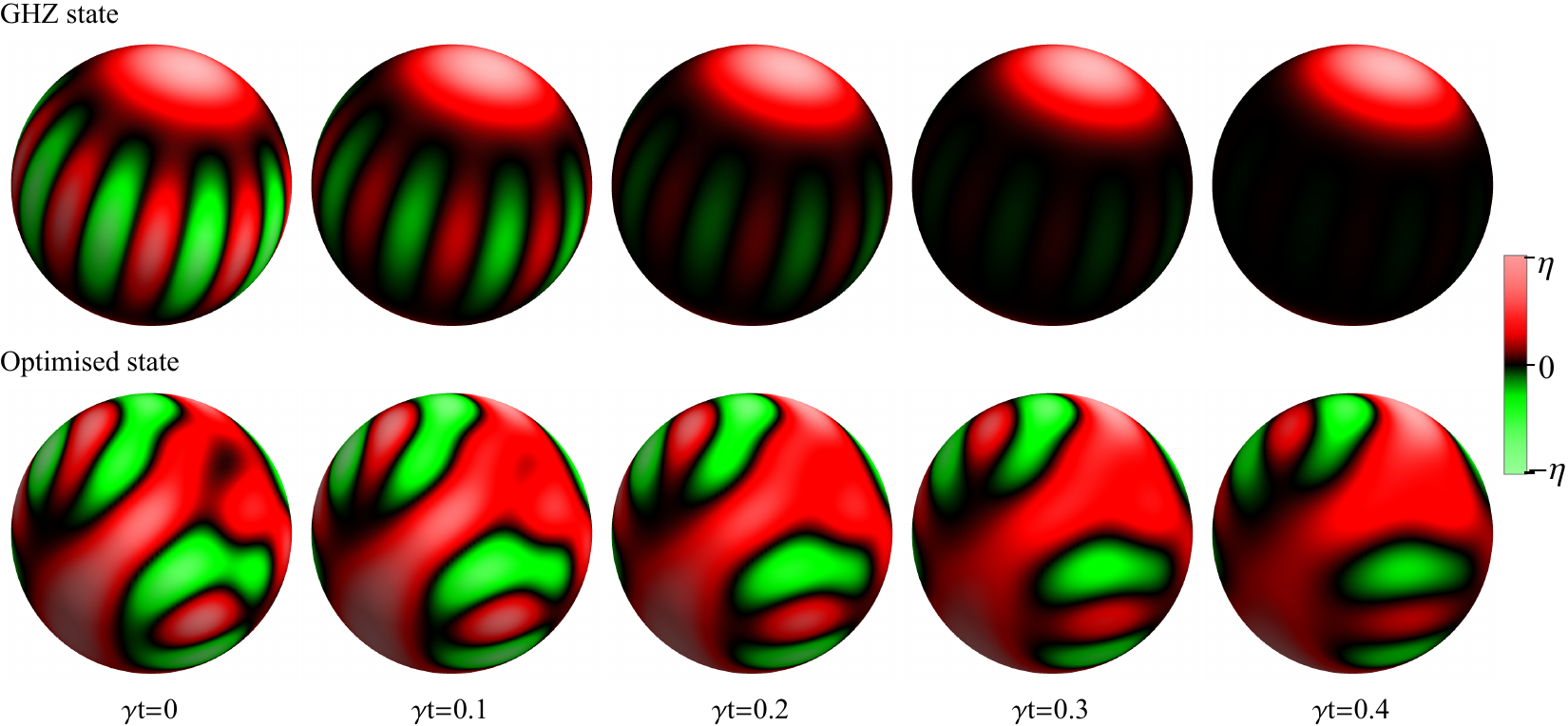}
		\caption{
			Wigner functions of permutation-symmetric $9$-qubit quantum states that evolve under dephasing noise.
			Time increases left-to-right and $\gamma t$ is the dimensionless time expressed in units
			of the decay time $\gamma^{-1}$.
			GHZ (upper) states are the most sensitive to an external magnetic field, but their coherences
			rapidly deteriorate due to fluctuations of the external field (as can be inferred from the rapidly fading
			coherences in their Winger functions). Our aim in the current work is to find states
			(lower) that are optimally sensitive to the external field while being robust against noise
			using variational techniques.
			These optimal states are not necessarily permutation symmetric (refer to Sec.~\ref{optNoise}).
			Red and green colours show positive and negative
			values of the function while brightness represents the absolute value of the function
			relative to its global maximum $\eta$.
			\label{introfig}
		}
	\end{centering}
\end{figure*}

Variational quantum circuits by construction depend only on a linear or polynomial number of parameters
while the Hilbert-space dimension of the underlying quantum state increases exponentially in the number of qubits.
This advantageous scaling allows one to tackle classically intractable problems.
The general concept of variational quantum algorithms is to prepare a parametrised quantum state using a
quantum processor and to vary its parameters externally until the optimum of a  suitable cost function is reached.
This cost function can be tailored to the particular problem.
For example, one can search for the ground state of a molecule by setting
the cost function to be the expectation value of the corresponding molecular Hamiltonian.
This technique is usually referred to as the variational quantum eigensolver
(VQE)~\cite{peruzzo2014variational,kandala2017hardware,moll2018quantum,mcclean2016theory}.
Quantum machine learning is another area where variational techniques may be valuable.
One is then interested in optimising a cost function that quantifies how similar the output
of the quantum circuit  is to a fixed dataset \cite{mitarai2018quantum,zhuang2019supervised}.
Moreover, it is also possible to recompile a quantum circuit into another by optimising a
metric on related quantum states \cite{jones2018quantum,heya2018variational}.

On the other hand, quantum metrology aims to enhance the precision
of a measurement process in the presence of statistical errors using entangled quantum states
\cite{giovannetti2004quantum,review,giovannetti11,toth14}.
For example, sensing magnetic fields with high precision is crucial in many applications,
such as determining chemical structure \cite{levittspind} or imaging living cells \cite{le2013optical}.
Various different types of high-performance magnetic field sensors have been developed,
including hall-effect sensors \cite{ramsden2011hall}, superconducting quantum interference devices
(SQUID)~\cite{huber2008gradiometric} and force sensors \cite{poggio2010force}.
In particular, in case of qubit-based magnetic field sensors, a qubit system interacts with
the magnetic field and the information about the magnetic field is encoded as an internal relative phase
of the quantum state. This information can then be extracted via a Ramsey-type measurement
\cite{bal2012ultrasensitive,wolf2015subpicotesla,ishikawa2012optical} that uses repeated projective
measurements. These experiments need to be iterated a number of times in order to
decrease the effect of statistical errors, such as the so-called shot noise.

If the probe state used in a metrology experiment
is an unentangled qubit state, the estimation
error of the external magnetic field (after a given number $\nu$ of fixed-duration field sampling experiments)
 is proportional to  $\nu^{-1/2}$. This scaling of the precision  is often referred to as
the standard quantum limit
(SQL)~\cite{giovannetti2004quantum,review,giovannetti11,toth14}
and can potentially be enhanced by using certain entangled states such as
GHZ states, symmetric Dicke states or squeezed states.
Although these entangled states offer a scaling of the estimation error beyond the standard quantum
limit $\nu^{-c}$  with $1/2 \leq c \leq 1$, they are also sensitive to noise
as illustrated in Fig.~\ref{introfig}.
In particular, it is well known that in the presence of uncorrelated Markovian dephasing,
the scaling achieved with a GHZ state is only the standard quantum limit \cite{ghzDephasing}.
It is our aim in the current work to derive quantum states that are robust to 
environmental noise but also sensitive to the external field of interest.
It is known from prior studies that quantum states subject to noise do not offer
an improved fundamental scaling \cite{channel_bounds,channel_bounds2,review,non-mark-noise,non-mark-noise2}
unless they are actively corrected during the environmental interaction 
\cite{error_corr1,error_corr2,error_corr3,error_corr4,error_corr5,error_corr6,errorcorr1,errorcorr2,errorcorr3}.
Nevertheless, optimisations of particular
probe states show that a significant improvement as a constant absolute factor
\cite{ghzDephasing,metrologyDecoherence,review} can be gained, even without active error correction.

Here, we propose a new variational quantum algorithm that optimises parametrised
probe states via a cost function that quantifies the metrological usefulness of a quantum state,
i.e., the precision of estimating the external field (refer to Sec.~\ref{secVarMetr}).
In particular, we envisage a quantum device that can generate a probe state
using a variational quantum circuit and this probe state can interact with the
external field of interest. We take into account the
simultaneous effect of decoherence due to environmental noise that deteriorates the
information about the external field contained in the probe state.
Finally, the output state is measured in a suitable measurement basis to
obtain the relevant cost function in a post-processing step, and the parameters
of the quantum circuit are updated accordingly.
This procedure is repeated until the optimal quantum state is found that achieves the
highest sensitivity under a given
noise model that is specific to the particular implementation of the device.
Note that available noisy quantum computer hardware, such as superconducting qubits,
can be straightforwardly used in this task either with or without error mitigation
or error correction.

We numerically simulate experiments under various different
error models in Sec~\ref{secNumSim} and comprehensively explore
systems consisting of up to $9$ qubits.
We find families of quantum states that non-trivially outperform 
previously known states.
In particular, our quantum algorithm searches within the exponentially large space
of non-symmetric states (as opposed to the linear dimensionality of symmetric ones)
and can therefore more fully exploit the potential offered by universal
quantum computers. We find that relaxing permutation symmetry
offers a significant improvement of the metrological sensitivity beyond 
symmetric states. 
This is a very interesting discovery and, perhaps, counter-intuitive as the problem of sensing
an external field under environmental noise (as in Eq.~\ref{noiseprocess}) is symmetric
under permutations and, to our best knowledge, states reported so far in the context of
quantum metrology are also symmetric \cite{ghzDephasing,dephaseOptState,dephaseOptState2,optstate1,optstate2,optstate3,optstate4}
e.g., GHZ, squeezed or symmetric Dicke states \cite{toth14,review}.
And note that those permutation symmetric states are indeed optimal in the noise-free scenario.
We develop an analytical understanding of a new family of highly entangled states that are not symmetric
under permutations but significantly outperform their symmetric analogues under experimental noise:
We derive an analytical model to show that these states can passively correct
first-order decay events of the amplitude damping channel and hence their superior
performance.
We emphasise that our aim in this work is to find practically relevant (near) optimal states
	of a small number of qubits $N$ that can be implemented in near term quantum
	hardware
	as opposed to finding asymptotically optimal states (i.e., when $N\rightarrow\infty$).

Furthermore, our algorithm is also practically useful for the following reasons.
First, in contrast to previous approaches which assumed and were limited to
specific noise models to maximise the metrological performance
\cite{dephaseOptState,dephaseOptState2,optstate1,optstate2,optstate3,optstate4},
our algorithm is independent of the choice of particular noise models. In particular,
we assume a fixed quantum hardware and our algorithm by construction finds the
metrologically optimal state that is tailored to the imperfections of that given device.
Moreover, our quantum measurement device can be implemented on
near-term quantum computers even without using quantum process tomography (which aims to
identify the dominant noise processes).
Our method is also quite general in view that the optimised parameters are not restricted
to circuit parameters: a number of generalisations
can be considered in that regard, e.g., incorporating the optimisation of an active error correction
\cite{errorcorr1,errorcorr2,errorcorr3},
and pulse control. Our scheme could also be extended to multiparameter estimation
quantum metrology. 

This manuscript is organised in the follwing way.
We begin by briefly reviewing key notions used in quantum metrology
in Sec.~\ref{quantumetsec}.
We then introduce the main idea of using variational
algorithms for quantum metrology in Sec.~\ref{secVarMetr}
and numerical simulations of these algorithms are
outlined in Sec.~\ref{secNumSim}. Our main results
on finding error-robust quantum states
are contained in Sec.~\ref{resultsSec}.
We finally outline an experimental setup
of our algorithm that could potentially be
implemented on
near-term hardware.

\section{Precision in quantum metrology\label{quantumetsec}}
We briefly recall basic notions used in quantum metrology in this section. We refer to reviews as, e.g., \cite{review,giovannetti11}, for more details.

Assume that the task is, e.g., to measure an external magnetic
field by using an initially prepared probe state $\ket{\psi}$ of $N$
qubits.
In this case the Hamiltonian in units of $\hbar = 1$
is proportional to the
collective angular momentum component $J_z$ as
\begin{equation*}
\mathcal{H} := \omega J_z = \omega \sum_{k=1}^N \sigma_z ^{(k)},
\end{equation*}
where $\sigma_z ^{(k)}$ is the Pauli $Z$ operator acting on
qubit $k$, and $\omega$ is the field strength to be probed.
If there are no imperfections, the time evolution of an
initially prepared probe state is described by the unitary operator
$U(\omega t)= \mathrm{exp}(-i t \omega J_z )$ that generates a global rotation of
all qubits.
One can subsequently perform projective measurements on identically prepared copies of
$|\psi(\omega t)\rangle := U(\omega t) |\psi\rangle$ and
results of these measurements can be used to estimate
the parameter $\omega$.
Note that if we take into account the effect of noise on
the system during the evolution period, the
 state to be measured is described by a density matrix $\rho_\omega$.

Let us assume that the measurement is described simply by an observable $O$ which decomposes into
the projectors
\begin{equation} \label{observable}
O = \sum_{n=1}^{d} \lambda_n |n \rangle \langle n |
\end{equation}
with $d=2^N$ and the expectation values of these projectors define
the probabilities of measurement outcomes $p( n |\omega ) := \tr[\rho_\omega \, |n \rangle \langle n |]$
that depend on the parameter $\omega$
\footnote{
Note that this simple measurement scheme generalises to POVM operators,
refer to \cite{nielsenChuang,review}
}.

\newcommand{\crb}{Cram\'er-Rao }
\newcommand{\psitheta}{| \psi(\underline{\theta}) \rangle}
\newcommand{\psinull}{| \psi_0 \rangle}
\newcommand{\maxprec}{(\Delta \omega)_{\mathrm{max}}^{-2}}

Measurements performed on $\nu$ identical copies of the state at a fixed $\omega$ can be used to
	estimate the value of $\omega$ via, e.g., a maximum likelihood estimator \cite{giovannetti11,toth14,review}.
The likelihood function 
tends to a Gaussian distribution \cite{giovannetti11,toth14,review}
for an increasing number of independent measurements $\nu$
that is centred at the true value $\omega$ and its inverse variance $\sigma^{-2}$ asymptotically
approaches the classical Fisher information $\nu F_c(O)$, which we will refer to as the precision.
In general, the estimation error $\Delta \omega$ of the parameter $\omega$ is bounded
by the so-called \crb  bound
\begin{equation}\label{crbound}
\Delta \omega \geq (\Delta \omega)_{\mathrm{CR}} := [ \nu  F_c(O) ]^{-1/2},
\end{equation}
where $F_c(O)$ is the classical Fisher information of the probability distribution
$p( n |\omega )$ that corresponds to eigenstates of the observable $O$ from Eq.~\eqref{observable}
and $\nu$ is the number of independent measurements.
The explicit form of the classical Fisher information can be specified in terms
of the measurement probabilities as
\begin{equation*}
F_c(O)  =  \sum_n p( n |\omega ) \bigg(\frac{\partial\mathrm{ln}    ~ p( n |\omega )}{\partial \omega} \bigg)^2.
\end{equation*}

The best possible estimation error using a fixed probe state can be obtained by maximising
Eq.~\ref{crbound} over all possible generalised measurements \cite{braunstein94,helstrom67,giovannetti11,toth14,review}
which leads to the so-called quantum \crb bound
\begin{equation}\label{qcrbound}
(\Delta \omega)_{\mathrm{CR}} \geq (\Delta \omega)_{\mathrm{max}} := [ \nu  F_Q(\rho_\omega) ]^{-1/2},
\end{equation}
where $F_Q(\rho_\omega)$ is the so-called quantum Fisher information of the state $\rho_\omega$
\cite{braunstein94,helstrom67,giovannetti11,toth14,review}.

This quantum Fisher information can be calculated for an arbitrary state $\rho_\omega$
via the expectation value $F_Q(\rho_\omega) = \tr[\rho_\omega L^2]$
of the Hermitian symmetric logarithmic derivative $L$ that is defined via
\begin{equation} \label{symmlogder}
 \frac{\partial \rho_\omega}{\partial \omega} := \frac{1}{2}(L \rho_\omega +\rho_\omega L).
\end{equation}
This symmetric logarithmic derivative can be obtained  for a density matrix
by first decomposing it into $\rho_\omega = \sum_{k} p_k |\psi_k \rangle \langle \psi_k | $
projectors onto its eigenstates $|\psi_k \rangle$ with $p_k>0$.
Matrix elements of the symmetric logarithmic derivative can then be obtained explicitly \cite{qfi1,qFisherComputation}
\begin{equation}\label{ll}
L_{ij} := \langle \psi_i | L | \psi_j \rangle = \frac{2}{p_i + p_j} \langle \psi_i | \frac{\partial \rho_\omega}{\partial \omega} | \psi_j \rangle.
\end{equation}
This formula simplifies for a unitary evolution as the derivative $\partial \rho_\omega/(\partial \omega)$
reduces to the commutator $i[\rho_\omega, \mathcal{H}]$.
Its calculation is more involved in case if the
evolution is not unitary \cite{qfi1,qFisherComputation}. Besides calculating the quantum Fisher information,
the symmetric logarithmic derivative is also useful for determining the optimal measurement basis.
In particular, performing measurements in the eigenbasis of $L$ saturates the quantum \crb bound
\cite{review,braunstein94}.

The fidelity between two density matrices \cite{nielsenChuang}
\begin{equation*}
\mathrm{Fid}(\rho_1,\rho_2) := (\tr[\sqrt{\sqrt{\rho_1} \rho_2 \sqrt{\rho_1}}])^2
\end{equation*}
is also related to the quantum Fisher information.
Assume that two density matrices $\rho(\omega)$ and $\rho(\omega + \delta_\omega)$
undergo the same noise process but one is exposed to an external field
$\omega$ while the other is exposed to $\omega + \delta_\omega$.
The quantum Fisher information in this case \cite{review} is recovered via the limit
\begin{equation} \label{fidelityrelation}
F_Q[\rho(\omega)] = 8 \lim_{\delta_\omega \rightarrow 0}  \frac{ 1-\mathrm{Fid}[\rho(\omega), \rho(\omega + \delta_\omega)] } { (\delta_\omega)^2}.
\end{equation}

\section{Variational state preparation for metrology \label{secVarMetr}}

\begin{figure}[tb]
	\begin{centering}
		\includegraphics[width=0.6\textwidth]{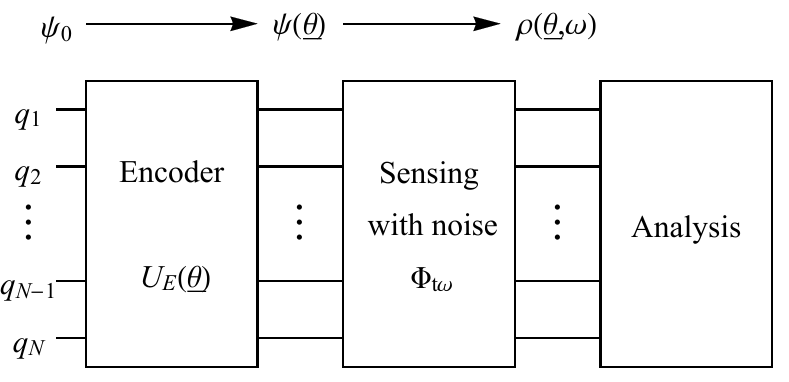}
		\caption{
			Circuit that potentially finds the quantum state $\psi(\underline{\theta})$
			that gives the best precision when estimating the parameter as the external
			field strength $\omega$.
			\label{device1}
		}
	\end{centering}
\end{figure}

We consider a hypothetical device that is depicted in Fig.~\ref{device1}
and which first initialises a system of $N$ qubits in the computational $0$ state.
Its parametrised encoder circuit then creates a probe state that is
exposed to the external field whose parameter $\omega$ is estimated.
The resulting state is finally analysed to
obtain an estimate of $\omega$.

	Let us assume that an optimal probe state $|\psi_{\mathrm{opt}}\rangle$
	maximises the metrological performance under a particular environmental noise model
	and experimental imperfections during the sensing period.
	We aim to \emph{approximate} this optimal state with a
	parametrised probe state $|\psi(\underline{\theta}_\mathrm{opt})\rangle \approx |\psi_{\mathrm{opt}}\rangle$ that is prepared
	via a shallow ansatz circuit $\psitheta:= U_E(\underline{\theta}) \psinull$
	which acts on the computational $0$ state $\psinull := | 0 \dots 00\rangle$
	of $N$ qubits.
	We assume for simplicity that this ansatz circuit is unitary and it decomposes into single
	and two-qubit quantum gates
	\begin{equation}\label{ansatzEq}
	U_E (\underline{\theta}) := U_\mu(\theta_\mu) \dots  U_2(\theta_2) U_1(\theta_1),
	\end{equation}
	each of which depends on a parameter $\theta_i$ with $i=\{1, 2, \dots \mu \}$.
	This parametrisation $\underline{\theta}$ corresponds to rotation angles
	of single and two-qubit quantum gates and it is externally optimised to find $\underline{\theta}_\mathrm{opt}$.
	We propose an explicit ansatz construction for the encoder as a shallow quantum circuit
	that depends on a classically tractable number of parameters
	in Sec.~\ref{probeStatesSec} and find that it can well approximate optimal states for various
	environmental noise models.

After preparing the ansatz state $\psitheta$ at a given set of parameters $\underline{\theta}$, it is exposed to the environment.
This process is characterised by a mapping $\Phi_{\omega t}(\cdot)$ of density matrices that
models the evolution under both the external field and under a non-unitary
noise process, and depends on both time $t$ and the parameter $\omega$.
We assume that this process is continuous in time.
Adapting results on infinitesimal divisible channels \cite{infdivchannel1,infdivchannel2},
we define the explicit action of this process on any density matrix $\rho$ via
\begin{equation} \label{noiseprocess}
\Phi_{\omega t}( \rho ) =
e^{-i  \omega t \mathcal{J}_z + \gamma t \mathcal{L}}  \rho ,
\end{equation}
where $\omega \mathcal{J}_z$ is the superoperator representation
of the external field Hamiltonian $\omega J_z := \omega \sum_{k=1}^N \sigma_z^{(k)} /2$
which generates unitary dynamics $\mathcal{J}_z \rho := [J_z, \rho]$
and the parameter $\omega$ is to be estimated.
The superopertor $\mathcal{L}$ generates non-unitary dynamics via a completely
positive trace preserving map between density operators and $\gamma$ is the decay
rate of the error model that can, in certain cases, depend on time
\footnote{In this work we consider a straightforward generalisation of Eq.~\eqref{noiseprocess}
	by allowing time-dependent decay rates as well $\gamma t \rightarrow f(t)$. These
	processes belong to the most general class of continuous channels, called infinitesimal divisible
	channels \cite{infdivchannel1,infdivchannel2}
}. Note that this form is very general and independent of
the particular choice of the noise model (as long as the process is continuous in time), 
and goes beyond previous investigations on quantum metrology \cite{metrologyDecoherence,dephaseOptState,dephaseOptState2}
that used noise models that commute with the external field evolution.

After exposing $\psitheta$ to the field, we denote its mixed state
by the density matrix $\rho(\omega t, \underline{\theta} )$. This state
contains information about the external field, which information is
deteriorated by noise during the evolution time $t$. The information
about the evolution can be read out during the analysis period.
By repeating the experiment $\nu$ times, the estimation precision of
the parameter $\omega$ can be increased. And this precision depends on the
amount of information about $\omega$ contained in the state
$\rho(\omega t, \underline{\theta} )$ and can be quantified using
the quantum Fisher information.
We aim to maximise this estimation precision 
by simultaneously varying the encoder parameters $\underline{\theta}$ and the exposure
time $t$ that the probe state spends in the noisy environment.
We thereby aim to simultaneously maximise the information
about the external field $\omega$ while minimising the effect of noise
on the probe state.
This will result in states $\psitheta$ that are (near)optimally sensitive to the external
field while being robust to noise.
We numerically simulate this procedure in
Sec.~\ref{secNumSim} and obtain (near) optimal states for metrology
using the estimation precision of $\omega$ as a target function while details
of an experimental implementation of our procedure have been deferred
to Sec.~\ref{secExpImp}. This experimental implementation has an explicit
construction of the analysis step and does not rely on the quantum Fisher information.

\section{Numerical simulations \label{secNumSim}}

We numerically (exactly) simulate the device introduced in the previous
section using the software package QuEST which can efficiently simulate
quantum circuits including noise processes \cite{quest}. We assume that the only source
of error is the evolution under the external field due to the process
$\Phi_{\omega t}( \cdot )$ and that the encoder and analysis steps
are perfect and require negligible time compared to the sensing time $t$.
These are considerably good approximations since the optimal sensing time $t$ is proportional
to the coherence time $1/\gamma$ (see below). This optimal time results in a significant buildup of error
during the sensing period independently of the decay rate $\gamma$ as also expected from, e.g., \cite{ghzDephasing,metrologyDecoherence}.
Moreover, the sensing time is significantly longer than the time
required by the encoder circuit and we assume the condition $\gamma \, t_\mathrm{enc} \ll 1$,
where $t_\mathrm{enc}$ is the time required for state preparation.
We remark that even if preparation is negligibly short, experimental imperfections of current and
near-term hardware, such as imperfections in quantum gates or measurement errors, might result in imperfect mixed
probe states and measurements. We model gate imperfections in \ref{withnoiseAppendix}
using state-of-the-art noise rates and demonstrate that the advantage of our
approach is still attainable when assuming imperfect preparation circuits.
We additionally assume that
each experiment can be repeated $\nu = T/t$ times, where $T$ is the overall
time of the metrology task.

We simulate a variety of encoder circuits that generate, e.g.,  GHZ, classical product
and squeezed states or arbitrary symmetric states. These states are introduced in more
detail in Sec.~\ref{probeStatesSec}.
After initialising the parametrised sensing state $\psitheta$, the evolution under
the external field is modelled using a Kraus-map representation of the process 
$\Phi_{\omega t}( \cdot )$ introduced in Eq.~\eqref{noiseprocess}. QuEST allows for modeling
arbitrary one and two-qubit errors \cite{quest} via their Kraus map representations
and we simulate various different error models in Sec.~\ref{optNoise} including, e.g.,
dephasing, amplitude damping and inhomogeneous Pauli errors.

The resulting density matrix $\rho(\omega t,\underline{\theta})$ could be used to
estimate the parameter $\omega$. As established in Sec.~\ref{quantumetsec}, the performance of this task is completely determined by the quantum
Fisher information $F_Q[ \rho(\omega t,\underline{\theta}) ]$ of this density matrix. Note that numerically
calculating the quantum Fisher information of $\rho(\omega t,\underline{\theta})$ 
avoids simulating the analysis step, however, it is completely equivalent to that.
We remark that explicit expressions are readily available in the literature \cite{qfi1,qFisherComputation}
	for computing the quantum Fisher information from the eigendecomposition of the density matrix.
	However, these formulas require the knowledge of the derivative $\partial_\omega \rho(\omega)$
	(or derivatives of eigenvalues and eigenvectors of $\rho$)
	which one might only be able to compute approximately using, e.g., a finite difference approach in $\omega$ --
	as in 	our explicit noise simulations using QuEST. 
	A finite difference in Eq.~\ref{fidelityrelation} might provide a
	numerically more stable and more efficient approximation than the aforementioned eigendecomposition of the approximate
	derivative of the density operator.

We calculate this quantum Fisher information and the resulting precision by
evaluating the circuit at two different evolutions.
Note that for noise channels which commute with the external field (see Sec.~\ref{optNoise})
the quantum Fisher information does not depend on the actual value of the parameter $\omega$.
However, in general the choice $\omega \rightarrow 0$ is adequate only when the field to
be sensed is sufficiently small. Nevertheless, we expect that most experimentally relevant environmental noise models approximately commute with the
external field evolution as discussed in Sec~\ref{optNoise}.
In our simulations we assume the density matrices $\rho_0 := \Phi_{0}( \psitheta )$
and $\rho_1 := \Phi_{\delta_\omega t}( \psitheta )$ via setting the parameter
in Eq.~\eqref{noiseprocess} as $\omega \rightarrow 0$ and $\omega \rightarrow \delta_\omega$,
respectively, and $\delta_\omega t \ll 1$.
We then approximate the precision via the fidelity
\begin{equation} \label{qfisherNumerical}
\maxprec
 = \tfrac{T}{t} F_Q[\rho_0] =
8 T  \frac{ 1-\mathrm{Fid}(\rho_0, \rho_1) } { t \, (\delta_\omega)^2} 
+ \mathcal{O}(\delta_\omega).
\end{equation}
Here we assume that the experiment can be repeated $\nu = T/t$ times, where
$t$ is the sensing time (approximately the overall time of executing the circuit once)
and $T$ is a constant (overall time of the metrology task). 
Note that the decay rate $\gamma$ from Eq.~\eqref{noiseprocess} 
is a parameter that can be set freely in the
simulations, however, the product $\gamma/T \maxprec$
is dimensionless and independent of both $\gamma$ and $T$,
refer to \ref{gammaIndep} and also to \cite{metrologyDecoherence}. 
We simulate metrology experiments with arbitrarily fixed $\gamma \gg \delta_\omega$
and optimise the dimensionless precision $\gamma/T \maxprec$
over the parameters $\underline{\theta}$ and $t$.
We finally obtain states that are (near) optimal for metrology in the presence of noise.

\section{Results \label{resultsSec}}

\subsection{Probe states \label{probeStatesSec}}
We simulate a variety of encoder circuits, but we do not aim to directly search in the
full, exponentially large state space of $N$ qubits. Note that this problem would
require encoder circuits that correspond to arbitrary unitary transformations
and would generally require exponentially many, i.e., at least $2^{N}$,
parameters to be optimised. Instead, we employ circuits that
contain a constant or linear number of parameters in the number of qubits
which can still sufficiently well approximate the optimal probe states $\psitheta$.
We also consider special cases of the general encoder circuit that generate, e.g., a family of squeezed states
or GHZ states, in order to compare our results to previously known states for metrology.

Results of the optimisations are shown in Fig.~\ref{prectime} for various error models and probe states.
In particular, probe states include GHZ (Fig.~\ref{prectime} red) and classical states (Fig.~\ref{prectime} black)
that we define as
\begin{eqnarray}
|\mathrm{GHZ}\rangle := \tfrac{1}{\sqrt{2}} |0\rangle^{\otimes N}  + e^{-i \phi}\tfrac{1}{\sqrt{2}} |1\rangle^{\otimes N},\\
|++\dots\rangle :=[ \tfrac{1}{\sqrt{2}} |0\rangle  + e^{-i \phi}\tfrac{1}{\sqrt{2}} |1\rangle ]^{\otimes N},
\end{eqnarray}
and their only parameters that we optimise are the phase angles
$\underline{\theta} = \phi$. Optimising these phase angles
improves the metrological performance in the case when 
noise is not rotationally symmetric around the external
field Hamiltonian, e.g., in the case of inhomogeneous Pauli errors.
We remark that GHZ states can be prepared using shallow circuits
consisting of a ladder of $N-1$ CNOT gates applied to the state $| + \rangle \otimes |  0 \rangle^{\otimes (N-1)}$.

One axis twisted squeezed states (Fig.~\ref{prectime} grey) are obtained \cite{spinSQ} by
the interaction under the permutation symmetric Hamiltonian
$J_z^2 = \sum_{k,l=1}^N \sigma_z^{(k)} \sigma_z^{(l)}$
and we define squeezed states via
\begin{equation*}
|\mathrm{sq}\rangle := e^{-i \theta_3 t J_z} e^{-i \theta_2 t J_x} e^{-i \theta_1 t J_z^2} 
[\tfrac{1}{\sqrt{2}} |0\rangle  + \tfrac{1}{\sqrt{2}} |1\rangle ]^{\otimes N},
\end{equation*}
and optimise their parameters $\underline{\theta} = (\theta_1,\theta_2,\theta_3)$.
Here $\theta_2$ generates a global rotation around the $x$ axis
to align the squeezing angle perpendicular to the external field Hamiltonian.
This unitary transformation can be represented by a quantum circuit
that contains parametrised controlled-$Z$ gates and parametrised local rotations
of the individual qubits \cite{dephaseOptState}. Data obtained for squeezed states
typically show an undulating trend in the number of qubits
throughout the graphs. This trend is due to the pairwise entanglement of squeezed
 states \cite{spinSQ}.

Optimised symmetric states (Fig.~\ref{prectime} brown) are obtained by a
direct search in the symmetric subspace whose dimension is linear in the
number of qubits. This subspace is spanned by so-called
symmetric Dicke states $|J=N/2,m\rangle$, where $N$ is the number of
qubits and $J$ is the total angular momentum and its $z$ projection is $m$,
refer to \cite{symmState,stockton2003,sakurai1995modern,schwinger65}.
Every symmetric state is then a linear combination of Dicke states
with complex coefficients $c_m$
\begin{eqnarray} \label{dickeStates}
|\mathrm{symm}\rangle := \sum_{m=-J}^J c_m |J=N/2,m\rangle.
\end{eqnarray} 
We optimise these coefficients in our algorithm
under the constraint that their absolute value
squares sum up to $1$ and $\underline{\theta} = \{c_m \} $.
Note that all states considered so far are symmetric
under permutations.

\begin{figure}[tb]
	\begin{centering}
		\includegraphics[width=0.7\textwidth]{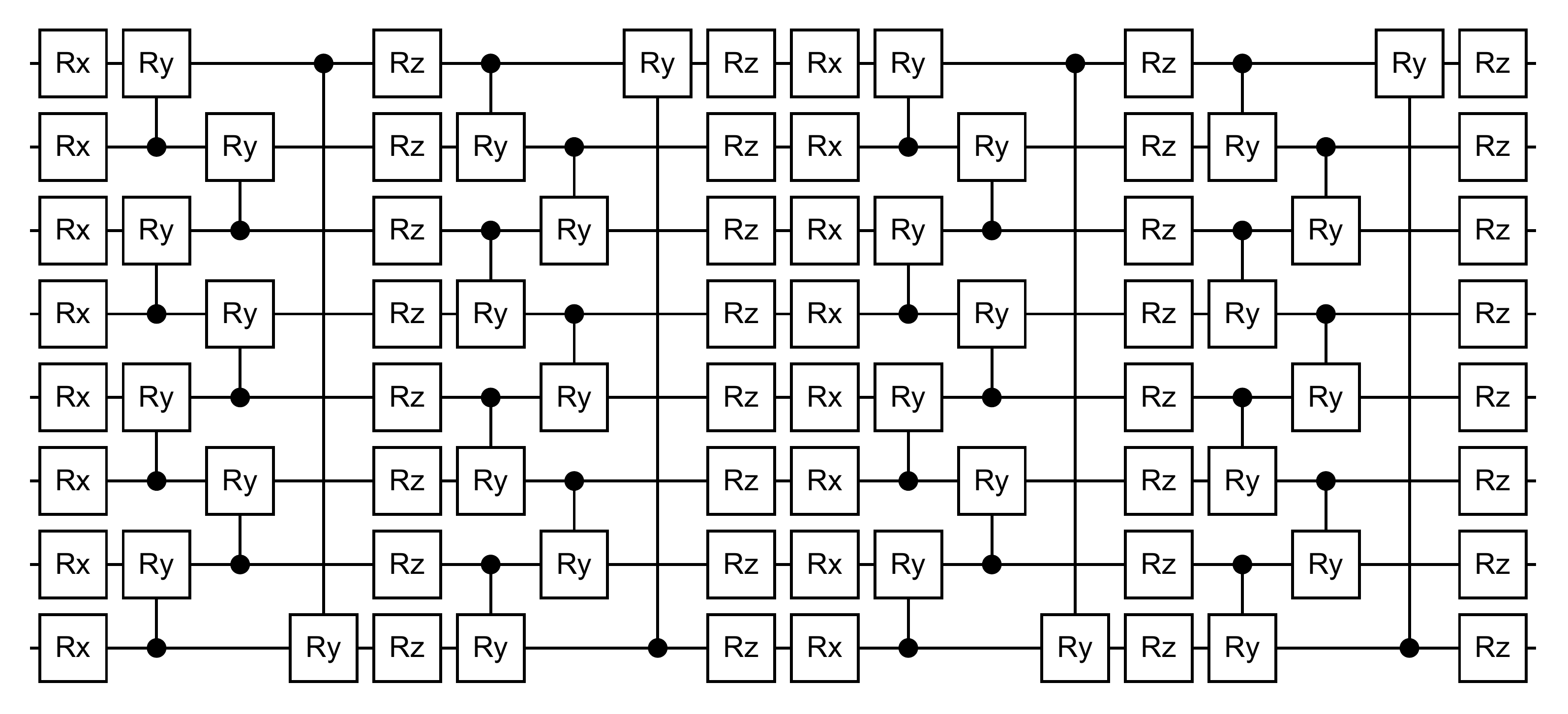}
		\caption{
			Example of the ansatz circuit for $N=8$ qubits.
			This circuit has a linear number of parameters in the
			number of qubits and can sufficiently well approximate
			states that are optimal for metrology under various
			different error models.
			\label{ansatz}
		}
	\end{centering}
\end{figure}

In contrast to the above introduced well-known symmetric states we aim to
search in the space of general qubit states via an ansatz circuit (Fig.~\ref{prectime} green)
that decomposes into single and two qubit gates from Eq.~\eqref{ansatzEq}.
In particular, we use a circuit shown in Fig.~\ref{ansatz} that has a classically tractable
number of parameters, i.e., linear in the number $N$ of qubits.
This ansatz structure is periodic and decomposes into repeated blocks.
The first block $B_1$ consists of $N$ single qubit $X$-rotations acting on individual qubits
with independent rotation angles $\theta_i$. The second block $B_2$ contains $N$ controlled-$Y$
rotations acting on nearest neighbour qubits followed by single qubit $Z$-rotations acting on individual qubits
and all gates have independent rotation angles $\theta_i$.
We found that the ansatz structure
$B_2 B_2 B_1 B_2 B_2 B_1$ can
well approximate metrologically optimal states for a variety of noise models for $N\leq9$,
even though  it cannot exactly reproduce arbitrary states in the exponentially large Hilbert space.
Moreover, increasing the number of repeated blocks
allows for better approximations of the optimal states.
We additionally remark that controlled-$Y$ rotations in our construction can easily be replaced by
specific hardware-native gates such as $XX$-gates.

\subsection{Optimisation of the ansatz parameters}
For the optimisation of the ansatz parameters, one might employ any suitable one of a range of methods: for example a direct search such as
Nelder–Mead (demonstrated experimentally in 2014\,\cite{peruzzo2014variational}),
or a systematic scan if the number of parameters is small~\cite{PRXH2}.
However, these might result in a slow convergence.
Stochastic gradient descent (see e.g. \cite{Rebentrost_2019})
was recently proven to converge
when assuming only a finite number of measurements for determining
the gradient vector \cite{sweke2019stochastic,kubler2019adaptive}.
This technique is expected to converge faster to
the optimum, however, a suitable approximation of the optimal
states as initial parameters are required.
We show in Sec.~\ref{analysisOfStates} that in case of a wide variety
of noise models optimal states are close to known ones, such as GHZ
or squeezed states, and we speculate that a good initial guess
can be attained in most practically relevant scenarios. 
We expect that the so-called quantum natural gradient approach \,\cite{koczor2019quantum}
might be superior to gradient descent as it corrects the gradient vector with
the natural metric in parameter space yet allows the optimisation of mixed quantum states and
non-unitary transformations
-- such as the time parameter $t$ in our optimisation.
Refer to Sec.~\ref{secExpImp}
for a possible experimental implementation of the optimisation procedure.

In our numerical simulations in Sec.~\ref{optNoise} we
consider relatively small systems as $N\leq 9$ and aim to
find the global	optimum without making any prior assumptions
on the optimal states. This allows us to also explore, e.g., non-symmetric states.
We used a randomised
adaptive coordinate descent search algorithm \cite{loshchilov2011adaptive} to find
the global optimum and made our code openly accessible \cite{gitcode}.
Although we cannot guarantee global optimality in general,
we compare our results in Sec.~\ref{optNoise} to analytically known optimal states in case of some noise
models and verify that indeed our ansatz-state optimisation can find 
a good approximation of the global optimum.
We additionally remark that our objective function is invariant under the permutation of any
qubits \emph{in general} and there exist $N!$ sets of
parameters that correspond to the same metrological performance in an $N$-qubit system.
This high degree of symmetry of the objective function can be exploited to speed
up the search even when no assumptions about the optimal parameters are made.

\subsection{Probe states optimised against noise \label{optNoise}}

\emph{Dephasing error ---}
It has been known that GHZ states perform equally well as classical
product states when undergoing dephasing \cite{ghzDephasing}, i.e.,
if the only source of noise is the stochastic fluctuation of
the parameter $\omega$ during the evolution period. We simulate metrology experiments
in the case when noise is dominated by dephasing. In
this special case all superoperators in Eq.~\eqref{noiseprocess}
commute and the evolution reduces to the explicit equation
\begin{equation}\label{dephaseChannel}
\Phi_{\omega t}( \rho ) =
[
\prod_{k=1}^N 
e^{ \gamma t \mathcal{L}_{\mathrm{de}}^{(k)} }  
] \,
e^{-i  \omega t \mathcal{J}_z}
\rho 
\end{equation}
which contains the superoperator $\omega\mathcal{J}_z$
that generates the unitary evolution under the external
field Hamiltonian $\omega J_z = \omega \sum_{k=1}^N \sigma_z^{(k)} /2$
and the non-unitary dephasing superoperator $\mathcal{L}_{\mathrm{de}}^{(k)}$
that effects all the $N$ qubits (indexed by $k$) identically and independently.
We use the Kraus map representation of the dephasing channel
that acts on an individual, single qubit via
\begin{equation}\label{dephaseDef}
e^{ \gamma t \mathcal{L}_{\mathrm{de}}^{(k)} }  \rho 
:=
[1-p(t)] \rho
+ p(t)  \sigma_z^{(k)} \rho \sigma_z^{(k)}
\end{equation}
and we define its time-dependent probability
as $p(t):= (1-e^{- \gamma t })/2$.
We apply this channel to the initialised probe state $\psitheta$
and calculate the dimensionless precision via the quantum fisher information of the
resulting density matrix $\rho(\omega t,\underline{\theta})$ as discussed below
Eq.~\eqref{qfisherNumerical}.

\begin{figure*}[tb]
	\begin{centering}
		\includegraphics[width=\linewidth]{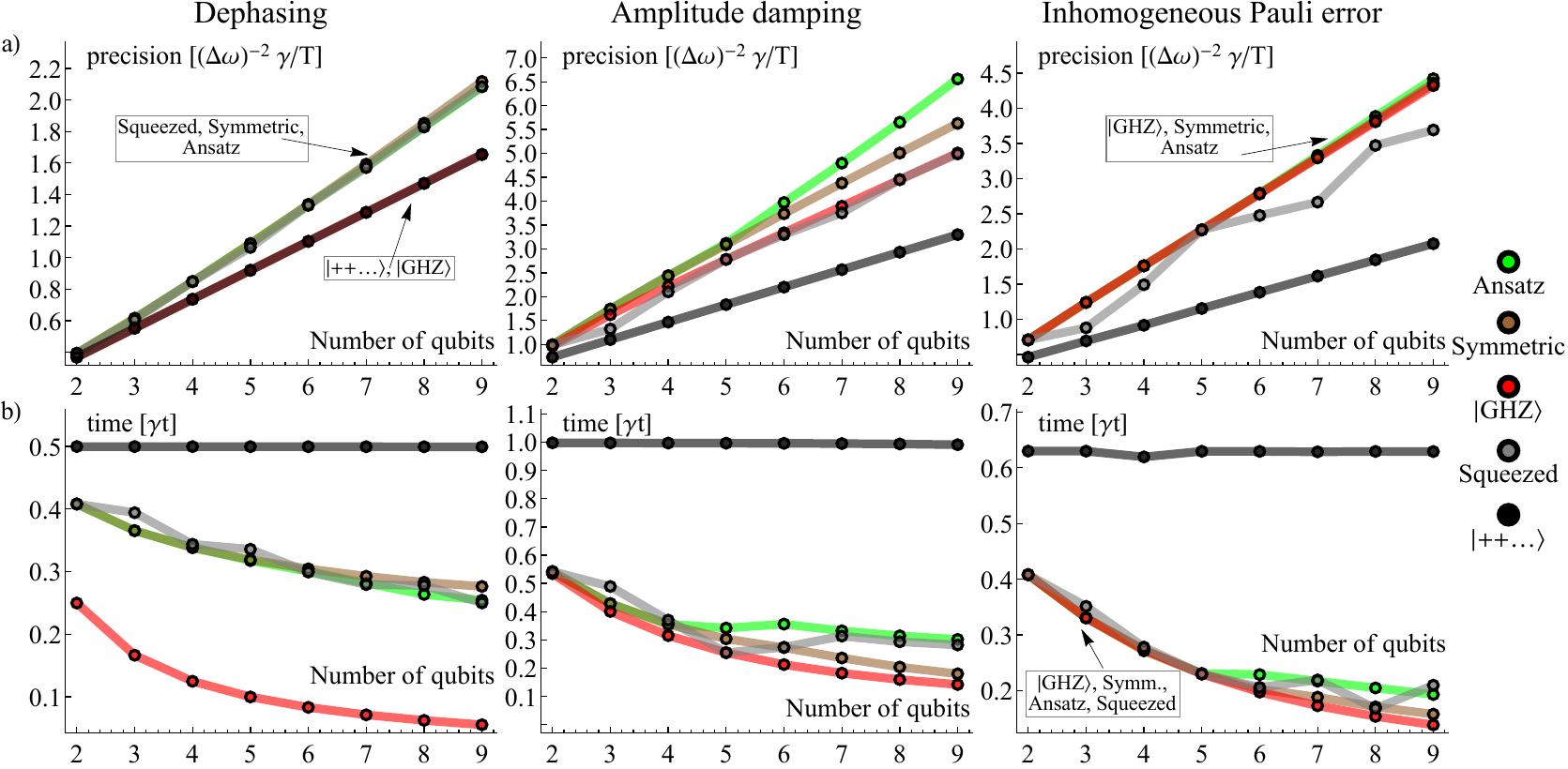}
		\caption{
			a) scaling of the optimised dimensionless precision as a function of the
			number of qubits calculated for a variety of probe states
			and noise models. Optimal ansatz states (green) obtained via
			the encoder circuit from Fig.~\ref{ansatz} outperform any symmetric
			state (brown) and break permutation symmetry in certain, practically relevant
			noise models.  
			Note that values on the $y$ axis are independent
			of the actual decay rate $\gamma$ of the noise model and independent
			of the overall time $T$ of the experiment (which consists of several repeated
			sub-experiments) when expressed in units of $\gamma/ T$.
			b) optimised probing time, i.e., optimal time that the probe
			state spends in the noisy environment. This time typically varies
			between $\propto 1 $ and $\propto 1/N$ in units of the error model's
			decay time and $N$ is the number of qubits.	
			\label{prectime}
		}
	\end{centering}
\end{figure*}

Fig.~\ref{prectime} a) (left) shows the scaling of the dimensionless
precision for a variety of different optimised probe states in case
of dephasing noise. The dimensionless precision of the previously discussed
GHZ and product states can be derived analytically as
$\gamma/T \maxprec = N / (2 e)$ where $e$
is the Euler number and Fig.~\ref{prectime} a) (left) GHZ (red) and separable (black)
states match the analytically derived formulas \cite{ghzDephasing}. 
This precision has a classical scaling, i.e., linear
in the number of qubits $N$.
Note that all states in Fig.~\ref{prectime} a) (left) display
a classical, linear scaling in the number of qubits which conforms with the asymptotic bounds on the quantum Fisher information obtained for usual Markovian channels
\cite{dephaseOptState,dephaseOptState2,fibre_bundle,channel_bounds,channel_bounds2}.
In particular, an upper bound on the quantum Fisher information
is saturated asymptotically by squeezed states \cite{ghzDephasing,dephaseOptState,dephaseOptState2}
in case of dephasing noise.
GHZ and separable states, therefore, can be outperformed by using optimised probe states but only up to an enhancement of a constant
factor at most $e \approx 2.72$
\cite{ghzDephasing,dephaseOptState,dephaseOptState2}.

The dimensionless precision achieved in our simulations
with squeezed states (grey) is nearly optimal and
results in a comparable performance to general symmetric states (brown) and
ansatz states (green) as also expected from \cite{ghzDephasing,dephaseOptState,dephaseOptState2}.
Our results conform with optimisations performed in \cite{ghzDephasing}
for a small number of qubits using symmetric states.
Note that our ansatz states (green) have a
negligible difference in performance when compared to general symmetric states (brown).
This difference is due to the fixed, finite depth of our ansatz circuit (which
can only \emph{approximate} arbitrary qubit states) and can be reduced by increasing the circuit depth.
Most importantly, this negligible difference in performance
	verifies that our approach using the ansatz circuit has (approximately) found the metrologically optimal states.

Fig.~\ref{prectime} b) (left) shows the optimal sensing times for the various 
probe states. These optimal sensing times can be derived analytically for the GHZ state
\cite{ghzDephasing} (in units of the decay time) as $\gamma t_\mathrm{opt} = (2N)^{-1}$
and for the classical product state as $\gamma t_\mathrm{opt} = 1/2$,
where $N$ is the number of qubits. The near-optimal squeezed (grey),
symmetric (brown) and general qubit (green) states tend to spend more time
than $(2N)^{-1}$ in the noisy environment
but less time than $1/2$.

\emph{Amplitude damping error ---}
We now consider a noise process in which amplitude damping or equivalently
spontaneous emission dominates. Similarly as with the dephasing channel,
all terms in Eq.~\eqref{noiseprocess} commute and the evolution reduces to
an analogous form with Eq.~\eqref{dephaseChannel}
but noise is now modelled using the damping superoperators
\begin{equation}
e^{ \gamma t \mathcal{L}_{\mathrm{da}}^{(k)} }  \rho 
:=
 K^{(k)}_1 \rho \, K^{(k)}_1
+  K^{(k)}_2 \rho \, [K^{(k)}_2]^\dagger
\end{equation}
that effect
all qubits identically and independently. 
We have used here the Kraus map representation of this channel
with the time-dependent Kraus operators
\begin{equation*}
K_1 :=
\left(
\begin{array}{cc}
1 & 0 \\
0 & \sqrt{1-p(t)} \\
\end{array}
\right),
\quad \quad
K_2 :=
\left(
\begin{array}{cc}
0 & \sqrt{p(t)} \\
0 & 0 \\
\end{array}
\right),
\end{equation*}
and their time-dependent probability is $p(t):= 1-e^{- \gamma t }$.

Fig.~\ref{prectime} a) (mid.) shows the scaling of the dimensionless
precision for a variety of different probe states that were optimised
against amplitude damping error. Note that all curves have a
linear, classical scaling in the number of qubits which conforms with the linear
asymptotic bound \cite{fibre_bundle,channel_bounds,channel_bounds2,dephaseOptState2}
on the quantum Fisher information obtained for this noise channel.
Under the amplitude damping error, GHZ
states (red) perform significantly better than classical product states (black).
Note that in the analysed region (2-9 qubits) squeezed states (grey)
closely approach the performance of GHZ states (red). On the other hand,
optimised general probe states offer significant improvements. In particular,
general symmetric states (brown) have a linear scaling but a steeper slope
than GHZ states. Moreover, relaxing permutation-symmetry constraints
on the probe state (green) results in further improvements. Although the
metrological task is permutation symmetric, i.e., its Hamiltonian and noise
model is invariant under permutations, our algorithm can discover non-symmetric states that
evidently outperform every symmetric state. These optimised ansatz states (green) are not
permutation symmetric for $N \geq 5$ and can apparently spend longer time in the environment.
Of course, the corresponding optimal measurement basis that saturates the \crb inequality consists of states
that are not permutation symmetric either.
Refer to Sec.~\ref{analysisOfStates} for a more detailed discussion of these states.
Moreover, we simulate state-of-the-art noise rates in \ref{withnoiseAppendix} and demonstrate
	that the superior performance of these non-symmetric states persists even
	when we take into account gate imperfections in their preparation circuits.

Fig.~\ref{prectime} b) (mid.) shows the optimal probing time. Optimised 
symmetric states (brown, grey) can spend more time in the noisy environment than GHZ states and
perform better. Ansatz states (green) have no permutation symmetry [Fig.~\ref{entdist} b) (mid.)]
for $N \geq 5$ and can apparently spend significantly more
time in the environment and this time appears to have a more preferable
scaling in the number of qubits when compared to symmetric states.
This advantage of ansatz states results in a significantly better performance
than optimised symmetric states. 
Refer to Sec.~\ref{analysisOfStates} for a more detailed analysis of the
resulting optimal states.

\emph{Inhomogeneous Pauli error ---}
The errors considered so far were rotationally symmetric with respect to the external field
and their superoperators therefore commute with the external field evolution.
In the following we consider an
error model that contains Pauli errors with no axial symmetry, such as bit flips.
In particular, we explicitly define and fix the process in Eq.~\eqref{noiseprocess}
at zero external field, i.e., $\omega = 0$ and at the particular time $\gamma t =1$
via the Kraus map 
\begin{equation} \label{inhomPauli}
e^{  \mathcal{L}_{\mathrm{pa}}^{(k)} }  \rho 
:=
[1-\sum_\alpha p_\alpha] \rho
+
\sum_\alpha \, p_\alpha  \, \sigma_\alpha^{(k)} \rho \sigma_\alpha^{(k)}
\end{equation}
that acts on each qubit identically and individually.
Here $\alpha \in \{x,y,z\}$ and the time-dependent probabilities
are asymmetric (inhomogeneous) $2 p_x =  p_y = 4 p_z$
and their sum is fixed to $\tfrac{3}{4} (1-e^{-1})$.
In the simulations we represent this Kraus map as a superoperator
matrix \cite{superoperator} whose matrix logarithm
then determines the generator $\mathcal{L}_{\mathrm{pa}}^{(k)} $.
The superoperator matrix of the entire process in Eq.~\eqref{noiseprocess}
is calculated via the matrix exponential of the sum
$-i  \omega t \sigma_z^{(k)}   + \gamma t \mathcal{L}_{\mathrm{pa}}^{(k)} $
for bounded time $0 \leq \gamma t \leq 1$.
This time-continuous process therefore interpolates between
the identity operation ($\gamma t=0$) and Eq.~\eqref{inhomPauli}
at the particular time $\gamma t = 1$.

Fig.~\ref{prectime} a) (right) shows the optimised dimensionless 
precision for a variety of probe states in case of inhomogeneous Pauli
errors. Note that GHZ states (red)
appear to be optimal, however, optimised probe states (grey, brown, green) tend to spend
more time in the environment than GHZ states for $N\geq 6$
as shown in Fig.~\ref{prectime} b). This results in a
slightly better performance of general symmetric states (brown).
Note that similarly to the amplitude damping channel, optimal ansatz states (green)
spend even longer time in the environment and outperform any symmetric state 
while breaking permutation-symmetry [Fig.~\ref{entdist} b) (right) for $N\geq 6$].
Although all curves display a classical, linear scaling, it is expected that
the steeper slope of ansatz states (green) results in higher
improvements for an increasing system size.

\begin{figure}[tb]
	\begin{centering}
		\includegraphics{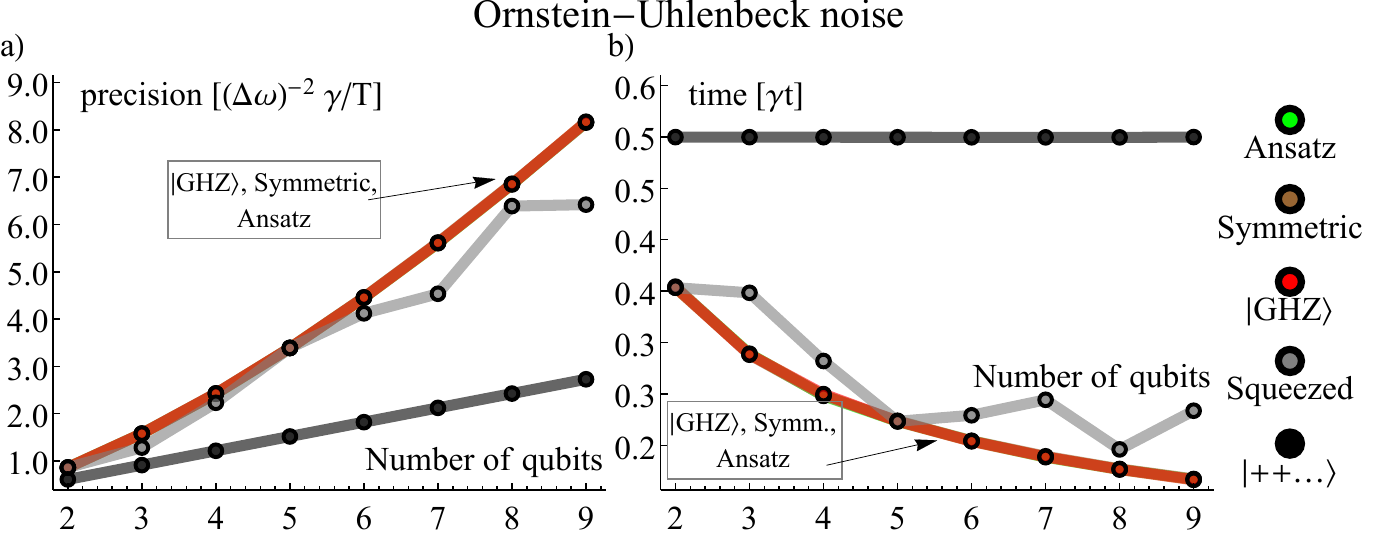}
		\caption{
			Optimised dimensionless precision a) and
			probing time b) for a variety of probe states in case if noise is dominated
			by a random fluctuation of the external field
			parameter $\omega$. We assume that this random fluctuation is described
			by the Ornstein-Uhlenbeck process in the limit of long correlation
			times, i.e., by a non-Markovian process. The zero-correlation-time limit
			yields dephasing from Fig.~\ref{prectime} (left).
			\label{timecorr}
		}
	\end{centering}
\end{figure}

\emph{Ornstein-Uhlenbeck noise ---}
We finally consider a noise model that is dominated by a random
fluctuation of the external field which follows the so-called
Ornstein-Uhlenbeck process \cite{ouprocess}. This noise process is
non-Markovian in general and in the limit of long correlation times
an improved scaling can be reached
when using GHZ states as already established by some of us in \cite{non-mark-noise}.
In particular, the time-dependent  fluctuation  of the external field is described
by its zero mean $\langle \omega'(t) \rangle = 0$ and a time-dependent correlation
function \cite{ouprocess,non-mark-noise0}
\begin{equation}
\langle \omega'(t) \omega'(\tau) \rangle = \tfrac{b \lambda}{2} e^{- \lambda |t-\tau| },
\end{equation}
where $\langle \cdot \rangle$ denotes the expected value, $\lambda^{-1}$ is
the process' finite correlation time and $b$ is the bandwidth of noise.
This process effectively results in a time-dependent buildup of a dephasing error
via the time-dependent probability $p(t) = [1 - e^{-f(t)}]/2$
and the noise channel is described by the Kraus-map representation
\begin{equation}
e^{ -f(t)  \mathcal{L}_{\mathrm{de}}^{(k)} }  \rho 
:=
[1-p(t)] \rho
+ p(t)  \sigma_z^{(k)} \rho \sigma_z^{(k)},
\end{equation}
from \ref{ouAppendix}. This
channel is analogous with simple dephasing from  Eq.~\eqref{dephaseDef}
up to the time-dependent decay rate $\gamma t \rightarrow f(t)$.
This time-dependent decay rate was derived explicitly in \cite{non-mark-noise0} 
as $f(t) := b [t + (e^{- \lambda t }-1)/\lambda] /2$.
Note that this noise model reduces to simple dephasing from
Eq.~\eqref{dephaseDef} in the limit of short correlation
times, i.e., when $b^{-1}\gg\lambda^{-1}$ and $\lambda^{-1}$ is the
correlation time. In  this case the decay rate
is characterised by $f(t) \approx b t/2$ the bandwidth of the classical
process \cite{non-mark-noise0}.

We now consider the limit of long correlation times with $b^{-1}\ll\lambda^{-1}$
as an example of non-Markovian channels.
In this case the time-dependent decay rate reduces to 
$f(t) \approx  b \lambda t^2 /4$ and  this process is analogous to
standard dephasing from Eq.~\eqref{dephaseDef} up to the time-dependent probabilities
$p(t) = [1 - e^{-(\gamma t)^2}]/2$ with $\gamma = \sqrt{b \lambda}/2$,
refer also to \cite{non-mark-noise0, non-mark-noise, non-mark-noise2}.
Fig.~\ref{timecorr} a) shows the optimised dimensionless precision for
various probe states.
 Note that the classical
product state can significantly be outperformed by using entangled quantum
states.
In particular, optimised probe states have an improved scaling, i.e., 
their dimensionless precision scales as $\maxprec \propto N^c$ in the number of qubits with $1<c\leq2$
as also expected from  \cite{non-mark-noise,non-mark-noise2}.
The increased, time-dependent buildup of noise forces the system to spend
the shortest possible time in the environment as shown in Fig.~\ref{timecorr} b).
GHZ states are therefore optimal and have an improved scaling
\cite{non-mark-noise,non-mark-noise2}. This again confirms that our approach (approximately) finds
	the optimal states.

\subsection{Analysis of the optimal states \label{analysisOfStates}}

\begin{figure*}[tb]
	\begin{centering}
		\includegraphics[width=\linewidth]{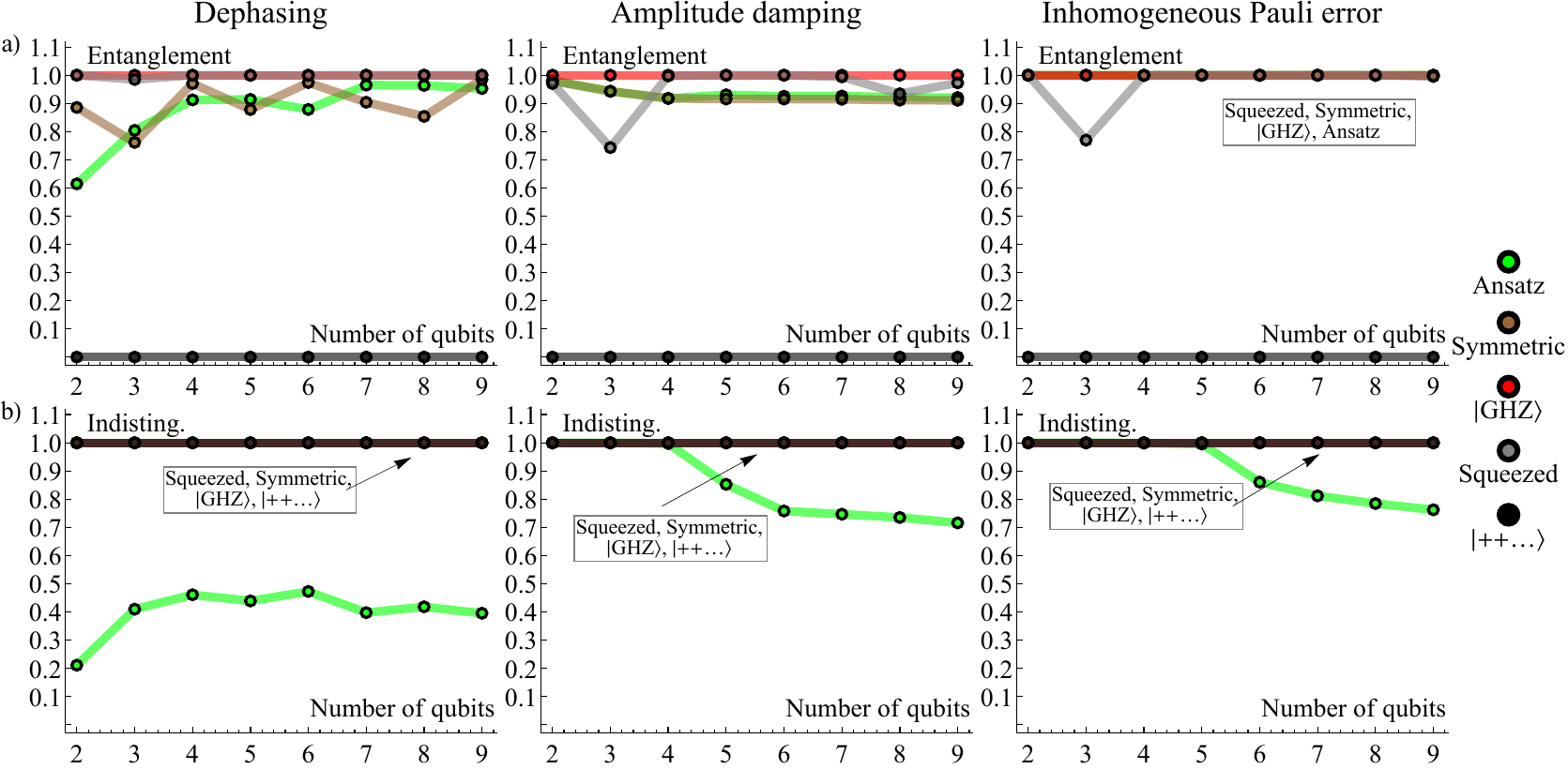}
		\caption{
			a) Linear entanglement of the optimised probe states which
			quantifies the average entanglement between a single qubit
			and the rest of the system, i.e., $N-1$ qubits. b) average
			indistinguishability of the qubits that form the optimal probe state.
			Only ansatz states can relax permutation symmetry, i.e., all other
			states can be expressed as linear combinations of Dicke states
			from Eq.~\eqref{dickeStates}. Relaxing this symmetry offers an
			improved metrological sensitivity of the optimised states, see Fig.~\ref{prectime}.
			\label{entdist}
		}
	\end{centering}
\end{figure*}

We analyse the optimised probe states by first calculating and plotting
simple measures that quantify their entanglement in Fig.~\ref{entdist} a)
and their permutation symmetry in Fig.~\ref{entdist} b).
In particular, we calculate an entanglement measure $S_{\mathrm{avg}}(|\psi\rangle)$ of the $N$-qubit
system via the average von Neumann entropy as
\begin{equation*}
S_{\mathrm{avg}}(|\psi\rangle) := \tfrac{1}{N} \sum_{k=1}^N - \tr[ \rho_k \log_2(\rho_k)],
\end{equation*}
where the single-qubit reduced density operator $\rho_k$ 
is obtained via the partial trace of the state $|\psi\rangle$ over all qubits except
qubit number $k$. This quantity is related to the Mayer-Wallach measure and quantifies the average
entanglement between a single qubit and the rest of the system, refer to \cite{mwmeasure1,mwmeasure2,mwmeasure3}.
Fig.~\ref{entdist} a) shows that classical product states (black) are unentangled
but all other optimised states are highly entangled.
It has been known that optimal states in case of dephasing 
are less entangled than GHZ states  \cite{ghzDephasing,dephaseOptState,dephaseOptState2}.
Optimised general symmetric (brown) and ansatz states (green) have slightly 
less linear entanglement than GHZ states in case of both dephasing and amplitude damping errors
Fig.~\ref{entdist} a) (left and mid.).
These optimised states are, however, very close to a GHZ state in case of  Pauli errors
in Fig.~\ref{entdist} a) (right) and have therefore similar linear entanglement.

We quantify permutation symmetry by calculating the 
the average fidelity of all permutations of the state $|\psi\rangle$
\begin{equation*}
P_{\mathrm{avg}}(|\psi\rangle) := \tfrac{1}{N_p} \sum_{k=1}^{N_p}
 \mathrm{Fid}[ |\psi\rangle , P_{k} |\psi\rangle ],
\end{equation*}
where $P_{k}$ permutes two qubits and $k$ runs over all distinct
permutations with $N_p = {N \choose 2}$ \footnote{and this measure can straightforwardly be
generalised to mixed states as well}.
Fig.~\ref{entdist} b) shows that all symmetric probe states have a maximal
permutation symmetry and only ansatz states (green) can relax this symmetry.
Optimal ansatz states Fig.~\ref{entdist} b) (green) clearly show a broken
permutation symmetry  in case of amplitude damping and inhomogeneous Pauli errors (mid. and right)
which results in a superior performance when compared to symmetric states.
Relaxing permutation symmetry offers a significant improvement
of the metrological sensitivity when noise is dominated 
by amplitude damping, c.f. green and brown lines in Fig.~\ref{prectime} a) (mid.).
Based on numerical evidence, we conjecture that
in the case of an even number of qubits the states $| \psi_a \rangle$ are near-optimal
against amplitude damping with suitable $c_1, c_2, c_3 \in \mathbb{C}$ 
\begin{equation} \label{optimalstate}
| \psi_a \rangle :=
c_1 |1 1 \cdots 1 \rangle 
+ c_2 | D \rangle 
+ c_3 |0 0 \cdots 0 \rangle,
\end{equation}
while the component $| D \rangle $ breaks permutation symmetry
and decomposes into the computational basis states
\begin{equation*}
\sqrt{\tfrac{2}{N}} ( |1 1 0 0 \cdots 0 0 0 \rangle + |0 0 1 1 \cdots 0 0 0 \rangle \cdots + |0 0 0 0 \cdots 0 1 1 \rangle ).
\end{equation*}
This additional component $| D \rangle $ in the state vector allows for individually resolving
first-order effects of the amplitude damping channel (i.e., flips of single qubits)
and results in an improved performance when compared to its symmetric counterpart,
the Dicke state $|J,J-2\rangle$ that decomposes into all permutations of a double excitation
with $J=N/2$.
For example, in case of $8$ qubits the explicit value of these coefficients is
$c_1 \approx 0.77$, $c_2 \approx 0.55$ and	$c_3 \approx 0.33$ while the resulting
metrological performance is close to $0.3 \%$ better than what we obtained with
the optimised ansatz circuit -- which can only approximate the state in Eq.~\eqref{optimalstate}.
Refer to \ref{symmAppendix}
for a fuller analysis of the superior power of these non-symmetric states.

\begin{figure*}[tb]
	\begin{centering}
		\includegraphics[width=\linewidth]{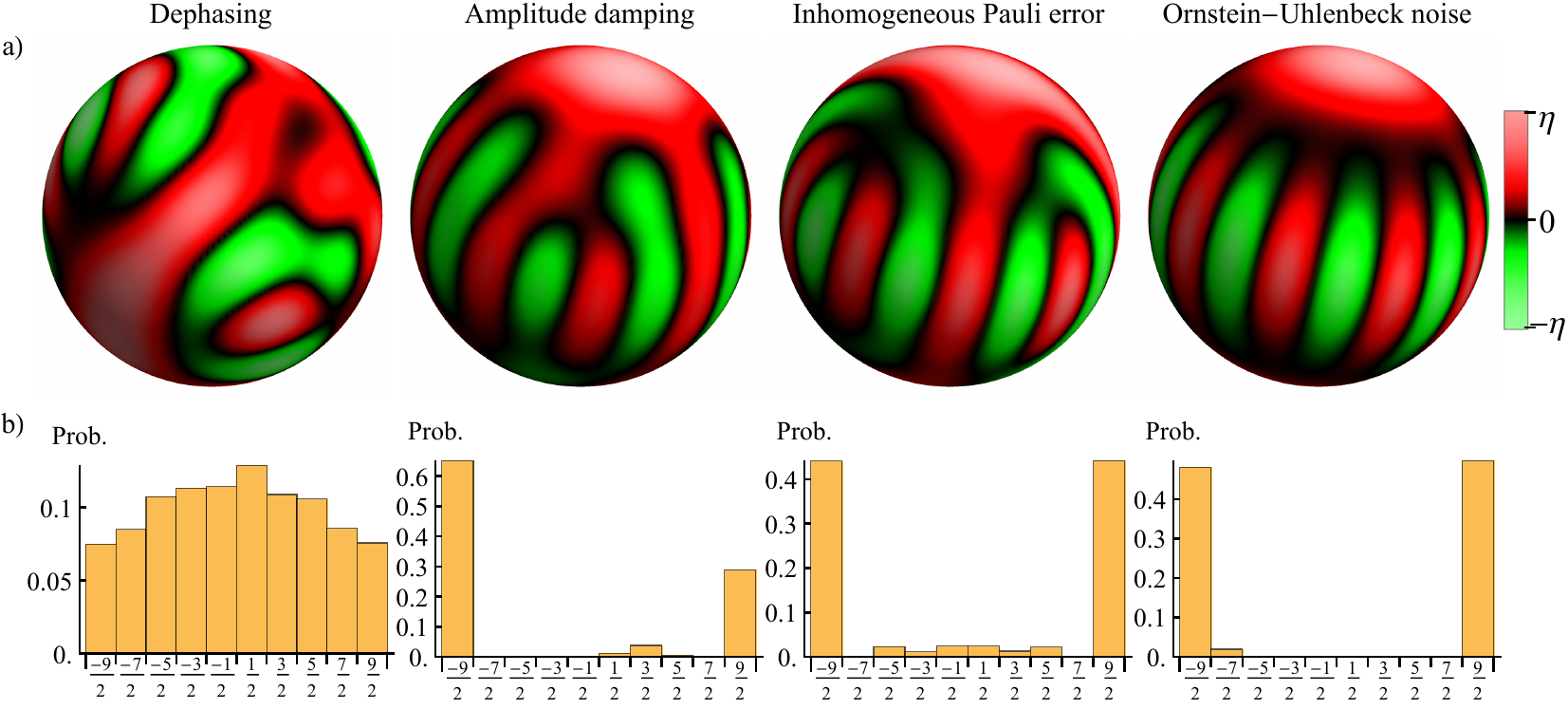}
		\caption{
			\label{wigner}
			a) Wigner functions of \emph{permutation symmetric} $9$-qubit states optimised against different
			error models from Fig.~\ref{prectime} a) (brown).
			The Wigner function in case of dephasing is similar to a squeezed state and
			in case of the Ornstein-Uhlenbeck process it is very close to a
			GHZ state. Wigner functions in case of amplitude damping and inhomogeneous Pauli
			errors are related to GHZ states. Red and green colours show positive and negative
			values of the function while brightness represents the absolute value of the function
			relative to its its global maximum $\eta$.
			b) Probabilities of Dicke states with $-9/2 \leq m \leq 9/2 $ as absolute value squares 
			of their optimised state-coefficients from Eq.~\eqref{dickeStates}.
		}
	\end{centering}
\end{figure*}

Symmetric states (brown) in Fig.~\ref{prectime} and in Fig.~\ref{timecorr}
are optimal in some error models (as in case of dephasing) and we analyse these states
separately. In particular, these state are linear combinations of Dicke states
from Eq.~\eqref{dickeStates} and probabilities of their optimised coefficients $c_m$
as $|c_m|^2$ are shown for $N=9$ qubits in Fig.~\ref{wigner} b).
Moreover, phase-space representations as the Wigner function offer an intuitive way for
visualising these permutation symmetric states. 
The Wigner function of an arbitrary mixed state is defined
as the expectation value
\begin{equation}
\label{inifnitedimdefinition}
W_\rho (\Omega) = \mathrm{Tr}\,[ \, \rho \, \mathcal{R}(\Omega)  \Pi_0  \mathcal{R}^\dagger(\Omega) ]
\end{equation}
of a rotated parity operator $\Pi_0$ where phase space is spanned by the
rotation angles $\Omega := (\theta, \phi)$ on the sphere 
and $\mathcal{R}(\Omega)$ is the rotation operator $\mathcal{R}(\Omega):= e^{i\phi \mathcal{J}_z} e^{i\theta \mathcal{J}_y} $,
refer to \cite{agarwal81,Brif98,koczor2017,KZG,koczor2016,thesis}
and to works \cite{Grossmann1976,koczor2018,tilma2016}
on rotated parity operators.
Fig.~\ref{wigner} a) shows Wigner functions of the optimal symmetric
states in case of $N=9$ qubits.

It has been known that squeezed states are optimal asymptotically
in case of dephasing \cite{ghzDephasing,dephaseOptState,dephaseOptState2}.
In our simulations, squeezed states are nearly optimal
in case of dephasing and Fig.~\ref{wigner} a) (left) shows typical characteristics of spin
(over)squeezed states. In particular, a squeezed Gaussian-like distribution
is surrounded by interference fringes. Moreover, Fig.~\ref{wigner} b) (left) identifies state-vector
coefficients that are related to squeezed states as the optimal symmetric state
consist of
a superposition of all Dicke states with a distribution of probabilities peaked at $m=0$.

GHZ states are optimal in case of the Ornstein-Uhlenbeck process
and Fig.~\ref{wigner} a) (right) clearly identifies the Wigner function of GHZ states
while Fig.~\ref{wigner} b) (right) shows an equal superposition of the spin-up and down states.

Symmetric states are suboptimal in case of amplitude damping and the best symmetric state
is similar to a GHZ state. In particular, it is a linear combination of the spin-up and down states
as shown in Fig.~\ref{wigner} b) (mid. left) but the state has a higher probability of being
in the spin-down state. Its Wigner function Fig.~\ref{wigner} a) (mid. left) is similar to 
Fig.~\ref{wigner} a) (right).

GHZ states are nearly optimal in case of  inhomogeneous Pauli errors
and Fig.~\ref{wigner} a) (mid. right) shows a Wigner function that is similar
to Fig.~\ref{wigner} a) (right).

\section{Possible experimental implementation \label{secExpImp}}

\begin{figure}[tb]
	\begin{centering}
		\includegraphics[width=0.6\textwidth]{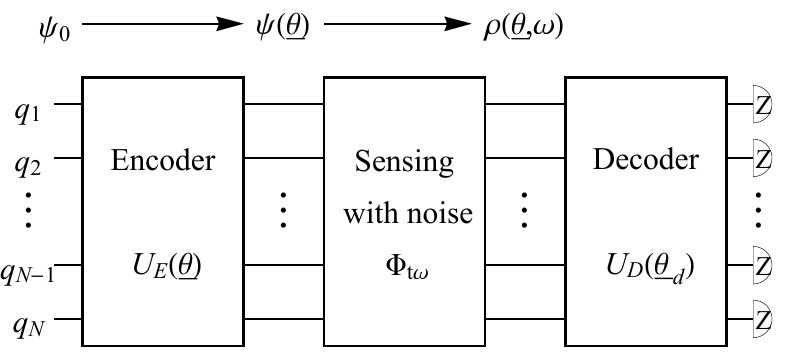}
		\caption{
			Circuit that potentially finds the quantum state $\psi(\underline{\theta})$
			that gives the best precision when estimating the parameter $\omega$ from
			projective measurements.
			\label{device2}
		}
	\end{centering}
\end{figure}

We now outline a possible experimental setup of our
approach  that can be implemented on near-term hardware.
In particular, we consider a hypothetical device depicted in Fig.~\ref{device2} which has a set of parameters
that can be varied externally.
This device can read out the evolution information
after the sensing period using a decoder circuit and a set of
projective measurements.
Results of $\nu$ repeated executions of this device
are used to estimate the precision of estimating $\omega$ and parameters of the encoder and decoder
circuits are variationally optimised to yield the best possible precision $\maxprec$.

Similarly as in Sec.~\ref{secVarMetr}, a probe state $\psitheta$
is initialised using an encoder circuit and this state is then
exposed to the noisy environment with the field to be probed.
The resulting mixed state $\rho(\omega t, \underline{\theta} )$ is now analysed using the
combination of a decoder circuit and a set of projective measurements in the computational basis.
In particular, a decoding circuit is applied to the state $\rho(\omega t, \underline{\theta} )$
that converts the evolution information $\omega t$ optimally
into probabilities of measuring the classical registers $|n\rangle$ 
at the end of the circuit. These classical registers are indexed using the binary numbers
$0 \leq n \leq 2^N-1$.
The measurement probabilities are given by the expectation values
\begin{equation}
p( n |\omega ) =  \langle n | \, U_D \rho(\omega t,\underline{\theta}) U_D^\dagger  \,  | n \rangle
\end{equation}
in the computational basis, i.e, in the eigenbasis of the collective Pauli $z$
operator $J_z := \sum_{k=1}^N \sigma_z^{(k)}$.
Note that the decoder circuit has the effect that it maps the computational basis states
$| n \rangle$ onto an arbitrary, effective basis $U_D^\dagger   | n \rangle$, therefore
mapping $J_z$ onto an effective observable $O(\underline{\theta}_d):= U_D^\dagger J_z U_D $.
This effective observable then
depends on the decoder parameters $\underline{\theta}_d$.

Recall that the classical Fisher information is related to the measurement probabilities in
the eigenbasis of the effective observable $O(\underline{\theta}_d)$
\begin{equation*}
(\Delta \omega)_{\mathrm{CR}}^{-2} = \nu F_c[ O(\underline{\theta}_d) ]  = 
\nu \sum_n p( n |\omega ) \bigg(\frac{\partial\mathrm{ln}    ~ p( n |\omega )}{\partial \omega} \bigg)^2
\end{equation*}
from Eq.~\eqref{crbound}.
The result of a single experiment using the above discussed setup yields a binary number $n$, the index of a
classical register into which the state has collapsed.
Repeating the experiment $\nu \gg 1$ times at the fixed $\omega$, the probabilities here can be estimated from the measurement results
and their derivatives can be approximated by repeating the
experiment at an external field $\omega + \delta_\omega$ and calculating
a finite difference. Our device can therefore estimate the precision $(\Delta \omega)_{\mathrm{CR}}$
that we aim to maximise over the decoder parameters $\underline{\theta}_d$.
\newcommand{\thvec}{\underline{\theta}}
\newcommand{\thd}{\underline{\theta}_d}

Note that this device has a set of parameters as $\thvec$, $\thd$ and $t$ that 
can be varied. In particular, maximising over the decoder parameters $\underline{\theta}_d$
optimises the observable, in the eigenbasis of which the measurements are effectively
performed. If the decoder and encoder circuits are universal, i.e.,
if $U_E(\underline{\theta})$ and $U_D(\underline{\theta}_d)$ span
the group $SU(2^N)$, then this setup can in principle achieve the combination
of an optimal sensing state $|\psi_\mathrm{opt}\rangle$ from Sec.~\ref{secVarMetr} and the corresponding
best measurement strategy.

Although the encoder and decoder circuits are not universal and not perfect 
in a practically relevant experimental implementation, we assume they can
approximate the precision via
\begin{equation}
 \nu \max_{\underline{\theta}_d} \, F_c[ O(\underline{\theta}_d) ]
\approx
 \nu  F_Q[ \rho(\omega t,\underline{\theta}) ]
= \maxprec ,
\end{equation}
that we calculated in the simulations
via the quantum Fisher information $F_Q[ \rho(\omega t,\underline{\theta}) ]$.
Note that the measurement process can be parallelised by
executing the task on several identical copies of the device.

Superconducting qubits are known to be excellent candidates for realising
both quantum computers and quantum sensors.
High-fidelity quantum gate operations and projective measurements
are a prerequisite for quantum computation and have
been successfully demonstrated in
\cite{barends2014superconducting,reed2012realization,kelly2015state}.
On the other hand, superconducting qubits can contain a SQUID-structure, and so the applied magnetic
fields can shift the resonant frequency
of the superconducting qubits~\cite{clarke2008superconducting}. There have been several experimental demonstrations
of using superconducting qubits as highly sensitive magnetic field sensors \cite{bal2012ultrasensitive,toida2019electron}.
Therefore, superconducting qubits are potentially suitable for experimentally demonstrating our proposal.

\section{Discussion and conclusion}
In this work we proposed variational quantum
algorithms for finding quantum states that
are optimal for quantum metrology in the
presence of environmental noise. Ours is not the first study to consider a classical optimisation of quantum states; for example Ref.~\cite{knott2016search} employs a
classical optimisation method to  obtain
metrologically useful states in
case of quantum optics.
This method is, however, limited to very small
quantum systems, i.e., 
when the average photon number is smaller than two
(due to the computational complexity of the problem).
Moreover, this approach does not take the
effect of noise into account. In the present study, we adapt state-of-the-art
variational techniques to tackle metrology in the presence of noise; moreover,
while the results we present so far have been obtained via classical simulations
(using the QuEST system) our technique can be operated on real quantum hardware
in order to explore beyond the classical reach.
Moreover, we provide an openly available \textsc{Mathematica} notebook in \cite{gitcode}
that contains our simulation code using QuEST and which can be used to reproduce
optimisation results contained in this manuscript as in Fig.~\ref{prectime}.

Our study has comprehensively
explored systems consisting of up to $9$ qubits:
we have numerically simulated experiments in case
of various different error models using the same ansatz-circuit structure.
We have demonstrated that our variational approach
using this fixed ansatz circuit is able to find previously known
optimal states despite vast differences in the error models.
Moreover, we found families of (near) optimal quantum states
that non-trivially outperform previously known symmetric states.
In particular, we demonstrated that relaxing
permutation symmetry of the probe states offers
significant improvements beyond symmetric states.
And we note that only these latter symmetric states have been
studied in detail so far in the context of quantum metrology
\cite{ghzDephasing,dephaseOptState,dephaseOptState2,optstate1,optstate2,optstate3,optstate4,toth14,review}.
We remark that verifying \emph{global} optimality of the symmetry-breaking
states is beyond the scope of the current work (although our numerical simulations
did (approximately) find the optimal states in case of the non-symmetry
breaking noise channels).

We analysed the resulting optimal states and found
that they are usually highly entangled but not
necessarily maximally entangled as can also be
expected from \cite{ghzDephasing}. We outlined
a possible experimental realisation that
could be implemented on near-term quantum hardware.

A number of natural extensions are apparent: we mention two examples here.
Firstly, the approach here can be extended to consider the case that the hardware used
to prepare the metrology state is itself noisy; our technique would then optimally use
such hardware, with-or-without the use of error mitigation techniques. Secondly, it
would clearly be interesting to combine the optimisation techniques mentioned here with
the error-detecting and error-correcting concepts described in, for example,
Refs.~\cite{error_corr1,error_corr2,error_corr3,error_corr4,error_corr5,error_corr6,errorcorr1,errorcorr2,errorcorr3}.

\ack
S.\,C.\,B. acknowledges financial support from the NQIT UK 
National Hub, EPSRC grant EP/M013243/1.
B.\,K. and S.\,C.\,B. acknowledge funding received from EU H2020-FETFLAG-03-2018 under the grant
agreement No 820495 (AQTION). 
S.\,E. acknowledges financial support from the Japan
Student Services Organization (JASSO) Student
Exchange Support Program (Graduate Scholarship for
Degree Seeking Students).
T.\,J. thanks the Clarendon Fund for their continued support.
Y.\,M. was supported by Leading Initiative for Excellent Young Researchers MEXT Japan, and was also supported by MEXT KAKENHI (Grant No. 15H05870).
B.K and S.E contributed to this work equally.
The authors are thankful to P. Zoller,
P. Silvi and R. Kaubruegger for useful comments
and their hospitality.

\appendix

\section{Deriving the dimensionless precision \label{gammaIndep}}

Recall that the precision is calculated using Eq.~\eqref{fidelityrelation}
via the fidelity
\begin{equation*}
\maxprec = \tfrac{T}{t} F_Q[\rho_0] =
8 T  \frac{ 1-\mathrm{Fid}(\rho_0, \rho_1) } { t \, (\delta_\omega)^2} 
+ \mathcal{O}(\delta_\omega),
\end{equation*}
between the density matrices $\rho_0$ and $\rho_1$.
In particular, the evolution process from Eq.~\eqref{noiseprocess} is set to $\omega \rightarrow 0$
for $\rho_0 := \Phi_{0}( \psitheta )$
 and to $\omega \rightarrow \delta_\omega$
and $\rho_1 := \Phi_{\delta_\omega t}( \psitheta )$
which results in the explicit form
\begin{eqnarray*}
\rho_0 & =  e^{ \gamma t \mathcal{L}}  |\psi(\underline{\theta}) \rangle \langle \psi(\underline{\theta}) | , \\
\rho_1 & =    e^{-i  \delta_\omega t \mathcal{J}_z + \gamma t \mathcal{L}}  |\psi(\underline{\theta}) \rangle \langle \psi(\underline{\theta}) | , 
\end{eqnarray*}
where $\psitheta$ is the probe state and $\mathcal{J}_z$, $\mathcal{L}$ are superoperators.
Let us now apply the transformation $t \rightarrow t' / \gamma$ and $\delta_\omega \rightarrow \delta_\omega'  \gamma$.
Note that this transformation does note effect the unitary
evolution, i.e., $\delta_\omega t = \delta_\omega' t'$, and results in the
density matrices
\begin{eqnarray*}
\rho_0' & =  e^{  t' \mathcal{L}}  |\psi(\underline{\theta}) \rangle \langle \psi(\underline{\theta}) | , \\
\rho_1' & =    e^{-i  \delta_\omega' t' \mathcal{J}_z + t' \mathcal{L}}  |\psi(\underline{\theta}) \rangle \langle \psi(\underline{\theta}) | , 
\end{eqnarray*}
which corresponds to the original dynamics but with effectively using a unit
decay rate $\gamma \rightarrow 1$.
The resulting precision therefore depends trivially on the parameter $\gamma$
\begin{equation*}
\maxprec =
8 T  \frac{ 1-\mathrm{Fid}(\rho_0', \rho_1') } { \gamma \, t' \, (\delta_\omega')^2} 
+ \mathcal{O}(\delta_\omega').
\end{equation*}
The precision is therefore a function $\maxprec = f(\gamma)$
of the decay rate  with $f(\gamma) = c/\gamma$ and the only degree of freedom
is the constant factor $c$. We finally obtain the dimensionless precision
$\gamma / T \maxprec $ that is independent of the decay rate
of our noise model.

\section{Kraus operators of the Ornstein-Uhlenbeck noise\label{ouAppendix}}

The Kraus representation of the Ornstein-Uhlenbeck noise has been derived in \cite{non-mark-noise0}
as in terms of the single-qubit Kraus operators
\begin{equation*}
K_1(t) :=
\left(
\begin{array}{cc}
q(t) & 0 \\
0 & 1 \\
\end{array}
\right),
\quad \quad
K_2(t) :=
\left(
\begin{array}{cc}
\sqrt{1-q^2(t)} & 0 \\
0 & 0 \\
\end{array}
\right),
\end{equation*}
with the time-dependent probability
$q(t) = e^{- f(t)}$ and $f(t) := \gamma[t + (e^{- \lambda t }-1)/\lambda]/2$.
It can be shown by a direct calculation that this Kraus map
is equivalent to simple dephasing 
\begin{equation*}
K_1(t) \rho K_1(t)
+
K_2(t) \rho K_2(t)
=
[1-p(t)] \rho
+ p(t)  \sigma_z^{(k)} \rho \sigma_z^{(k)}
\end{equation*}
up to the time-dependent probability of dephasing $p(t) = [1 - e^{-f(t)}]/2$.

\section{Improved performance of non-symmetric states\label{symmAppendix}}
Let us define the Dicke states $|J, J \rangle := |0 0 \cdots 0 \rangle $
and $|J, -J \rangle := |1 1 \cdots 1 \rangle $ with $J=N/2$
for an even number of qubits $N$.
By analysing the optimal states produced by the variational technique,
i.e. the upper line in Fig.~\ref{prectime}(mid), one can observe that
for even numbers of qubits the states are close to $| \psi_a \rangle$ given by
\begin{equation*}
| \psi_a \rangle :=
c_1 |J, J \rangle 
+ c_2 | D \rangle 
+ c_3 |J, -J \rangle,
\end{equation*}
where the component $| D \rangle$ breaks permutation symmetry
and its explicit form is
\begin{equation*}
\sqrt{\tfrac{2}{N}}( |1 1 0 0 \cdots 0 0 0 \rangle + |0 0 1 1 \cdots 0 0 0 \rangle \cdots + |0 0 0 0 \cdots 0 1 1 \rangle ).
\end{equation*}
For $N=6$ and $N=8$ the performance of $| \psi_a \rangle$ almost exactly
coincides with that of the variationally-determined states.
In the following, we compare the states $| \psi_a \rangle$ to their symmetric analogues
\begin{equation*}
| \psi_s \rangle :=
c_1 |J, J \rangle 
+ c_2 |J, J-2 \rangle
+ c_3 |J, -J \rangle. 
\end{equation*}
The latter under-perform the former, symmetry-broken states by a
considerable margin (see Fig.~\ref{prectime}(mid)) but are 
nearly optimal {\em within} the restriction to symmetric states
(performance $> 99\%$ of the
optimal symmetric state for $N=8$).  Here the Dicke
state $|J, J-2 \rangle$ contains all distinct $\binom{N}{2}$ permutations
of a double excitation.

In the following, we provide an analysis into the reason that these non-symmetric
states have superior performance. Roughly speaking, the explanation is that the basis
of meaningful measurements is larger, with measurements corresponding to specific
first-order decay events. This increased basis leads to a correspondingly greater
Fisher information.

After exposing these states to the noisy environment, their density
matrices $\rho_a$ and $\rho_s$ contain information about the external field.
This information is extracted by applying a decoder circuit to these states
and performing measurements in the computational basis. The decoder
circuit acts as, e.g.,  $U_d \rho_a (U_d)^\dagger$ and the probability of
measuring a computational basis state $|n \rangle$ is given by the expectation
value  $\langle n  |U_d \rho_a (U_d)^\dagger |n \rangle $.
This probability can be formally written as measuring in an alternative
basis $|n' \rangle = (U_d)^\dagger |n \rangle $ without the application
of a decoder circuit and the equivalent probability is then given by
 $\langle n'  | \rho_a |n' \rangle $. We will
refer to it as the optimal measurement basis.

\newcommand{\tplus}{T_{+}}
If there is no noise
present (i.e., $\gamma=0$) the optimal decoder circuit results for both $\rho_a$ and $\rho_s$
in the optimal measurement basis as $b_1 |J, J \rangle  \pm b_2 |J, -J \rangle$
with $b_1= e^{i \pi/2} b_2 = 1/\sqrt{2}$ and all other basis states are found
with exactly $0$ probability
\footnote{These states with  $|c_2|>0$ are not optimal in the noiseless case 
and result in non-zero probabilities of bases that contain
double excitations, but this does not effect our discussion}.
In the case of finite noise (i.e., $\gamma >0$)
$T_1$ relaxation events will occur but only to individual qubits in the first order.
This can be shown by decomposing the Kraus operators of the amplitude damping channel
into an identity operation plus $T_1$ and $T_2$ relaxation terms.
In particular, the operator $T_{+} := | 0\rangle \langle 1 |$ is denoted as $\tplus^{(j)}$
when acting on qubit $j$ and results in, e.g., $\tplus^{(1)} |111 \cdots\rangle =  |011 \cdots\rangle$.
These events will result in non-zero probabilities of additional measurement results.
In particular, the optimal measurement basis of the symmetric case $\rho_s$ will include
the basis $|S\rangle:= b_1 |J, J-1 \rangle  \pm b_2 |J, -J +1\rangle$ which has a non-zero
probability $\mathrm{prob}(S) :=  \langle S|\rho_s|S\rangle$ only because of the first-order effect of $T_1$ relaxation.

In case of the non-symmetric state $\rho_a$, the optimal
measurement basis includes all $2N$ bases 
\begin{equation}
|A_j\rangle :=\tplus^{(j)} ( b_1 |J, -J \rangle   \pm    b_2 \sqrt{N/2} | D \rangle)
\end{equation}
which contain flips of individual qubits $1 \leq j \leq N$ via the operator $\tplus$.
For example, flipping the first qubit yields
\begin{equation*}
|A_1\rangle
= 
b_1 |0 1 1 1 \cdots 1 1 1 \rangle \pm b_2 |0 1 0 0 \cdots 0 0 0 \rangle,
\end{equation*}
or flipping the second yields
\begin{equation*}
|A_2\rangle
= 
b_1 |1 0 1 1 \cdots 1 1 1 \rangle \pm b_2 |1 0 0 0 \cdots 0 0 0 \rangle,
\end{equation*}
and similarly
\begin{equation*}
|A_3\rangle
= 
b_1 |1 1 0 1 \cdots 1 1 1 \rangle \pm b_2 |0 0 0 1 \cdots 0 0 0 \rangle.
\end{equation*}
There are overall $2N$ of these measurement bases and their non-zero
probabilities $ \langle A_j|\rho_a|A_j\rangle$ are only due to
the effect of first-order $T_1$ relaxation.

In the following, we aim to comparing the classical Fisher information
due to these additional bases that extract information about the
external filed even after the state has undergone relaxation.
In particular, its contribution to the classical Fisher information
in the symmetric scenario is given by
\begin{equation}\label{fisher1}
F_s = \mathrm{prob}(S) ^{-1} \bigg(\frac{\partial \mathrm{prob}(S) }{\partial \omega} \bigg)^2.
\end{equation}
with the probability defined as $\mathrm{prob}(S) :=  \langle S|\rho_s|S\rangle$.
Similarly for the non-symmetric case the contribution to the classical Fisher information
is given by the sum
\begin{equation}\label{fisher2}
F_a = \sum_{j=1}^{2N} \mathrm{prob}(A_j) ^{-1} \bigg(\frac{\partial \mathrm{prob}(A_j) }{\partial \omega} \bigg)^2
\end{equation}
over all $2N$ bases with $\mathrm{prob}(A_j) :=  \langle A_j|\rho_s|A_j \rangle$.

\begin{figure}[tb]
	\begin{centering}
		\includegraphics{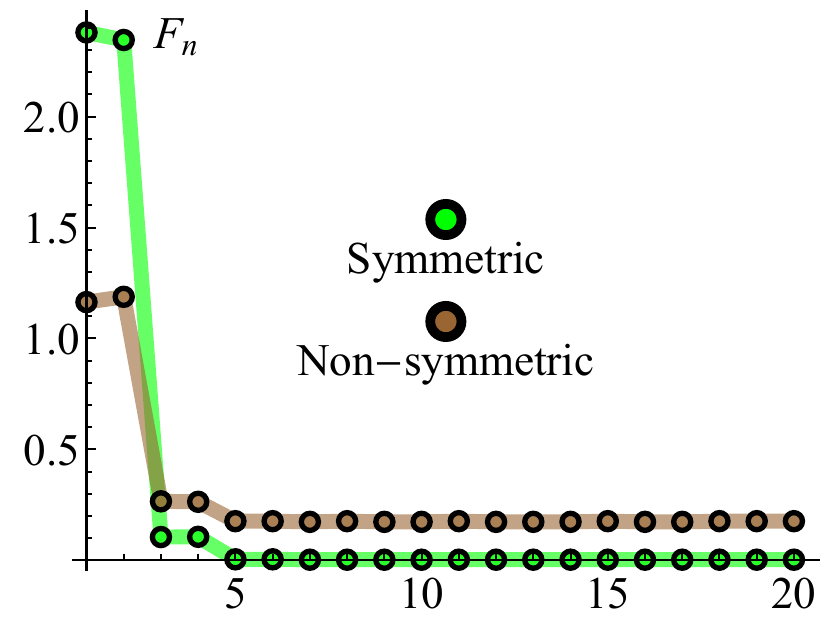}
		\caption{
			Contribution of the individual measurement bases to the classical Fisher
			information $F_n$, i.e., the sum of these individual elements results in the quantum Fisher
			information. (green) optimal measurement scheme with the optimised symmetric input state $\psi_s$ and (brown)
			with the optimised non-symmetric input state $\psi_a$ for $N=8$ qubits. The last 16 orange dots
			are due to measurements in the bases $|A_j\rangle$.
			The optimal symmetric state is close to a tilted GHZ state, i.e., $c_2 \approx 0$ and the first
			two green dots are therefore higher than in case of the non-symmetric scheme (brown) where $c_2 \approx c_1$.
			The optimised non-symmetric state is a trade-off between increasing the contribution $c_2 |D\rangle$
			(which increases the last 16 brown dots)
			at the cost of decreasing the absolute value of $c_1$ and $c_3$
			(which decreases values of the first two brown dots).
			\label{nonsymm}
		}
	\end{centering}
\end{figure}

In the symmetric scenario, these measurement probabilities are calculated 
including the effect of external field evolution as
$\mathrm{prob}(S)  = \sum_{k=1}^N |\langle S|  \tplus^{(k)} e^{i \omega t J_z} | \psi_s \rangle|^2$
and assuming that a relaxation event occurred to qubit $k$. The overlaps can be calculated
explicitly
\begin{equation*}
\langle S|  \tplus^{(k)} e^{i \omega t J_z} | \psi_s \rangle
= e^{i \omega t N} c_1/\sqrt{N}
+ e^{i \omega t 2} c_2 \frac{\sqrt{N-1}}{\sqrt{N^2/2}},
\end{equation*}
using the overlaps $\langle J, -J+1 | \tplus^{(k)} | J,-J \rangle = 1/\sqrt{N}$
and similarly $\langle J, J-1 | \tplus^{(k)} | J,J-2 \rangle = (N-1)/\sqrt{N} \binom{N}{2}^{-1/2}$.

The probabilities in the  non-symmetric scenario are similarly
 $\mathrm{prob}(A_j)  = \sum_{k=1}^N  |\langle A_j|  \tplus^{(k)} e^{i \omega t J_z} | \psi_a \rangle|^2$
 given by the overlaps which we calculate explicitly as
\begin{equation*}
\langle A_j|  \tplus^{(k)} e^{i \omega t J_z} | \psi_a \rangle
= \delta_{kj} (e^{i \omega t N} c_1
+   \, e^{i \omega t 2} c_2 /\sqrt{N/2}),
\end{equation*}
where $\delta_{kj}$ is the Kronecker delta.
Summing up for all qubits indexed by $k$, we obtain
the ratio of probabilities
\begin{equation}
\mathrm{prob}(S) / \mathrm{prob}(A_j) = \mathcal{O}( N^0).
\end{equation}
Partial derivatives of the probabilities have a similar ratio
\begin{equation}
\frac{\partial \mathrm{prob}(S) }{\partial \omega}
\left[\frac{\partial \mathrm{prob}(A_j) }{\partial \omega}\right]^{-1}
= \mathcal{O}( N^0).
\end{equation}
The measurement in the non-symmetric scheme is as
sensitive as in the symmetric one. However, in the
non-symmetric scheme there are $N$ distinguishable measurement
bases $|A_j\rangle $.
These probabilities $\mathrm{prob}(A_j) $ and their derivatives determine the classical Fisher information
via Eq.~\eqref{fisher1} and via Eq.~\eqref{fisher2}.
The increasing number of measurement bases in the non-symmetric scheme therefore results in a superior scaling $\propto N$
of the corresponding classical Fisher information (when compared to the symmetric scenario) due to the
ability of measuring in the $2N$ distinct bases $|A_j\rangle $, refer to Fig.~\ref{nonsymm}.

In conclusion, additional information via the above discussed scheme is extracted only
from \emph{that part} of the state to which amplitude damping occurred.
It can be viewed as a correction (albeit a significant one) to the zeroth order part which both the symmetric
and non-symmetric cases share. And this correction is even comparable to the zeroth order part in case
of the non-symmetric scheme.

\section{Effect of non-ideal quantum circuits\label{withnoiseAppendix}}

\begin{figure}[tb]
	\begin{centering}
		\includegraphics[width=0.8\textwidth]{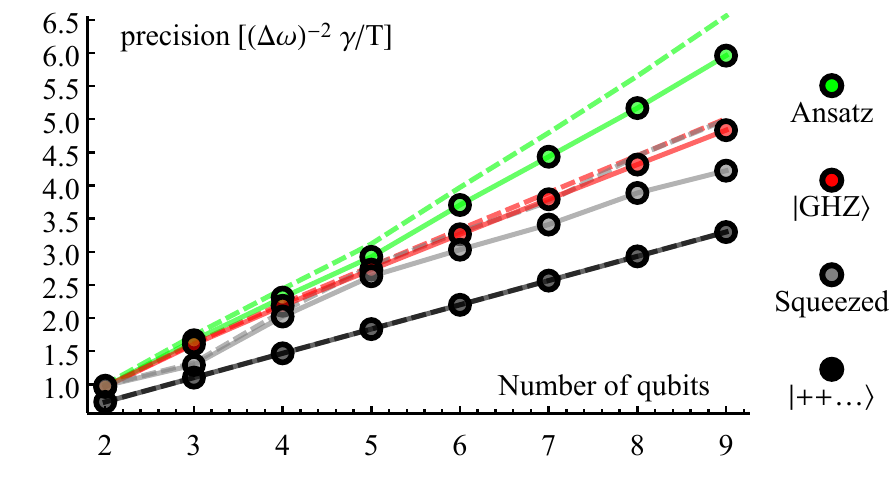}	
		\caption{
			Performance of the different families of quantum states under amplitude damping
			when their preparation circuits are imperfect -- individual gates undergo
			depolarising noise with a probability $10^{-3}$ in case of two-qubit
			gates and $10^{-4}$ in case of single-qubit gates. The superior performance
			of ansatz states over previously known ones persists despite their
			deeper preparation circuits. Dashed lines show the performance attained with
			perfect, unitary preparation circuits.
			\label{withnoise}
		}
	\end{centering}
\end{figure}

We have repeated the numerical experiments of Sec.~\ref{optNoise} with taking imperfections
of the preparation circuits into account. We do not consider here the family of arbitrary symmetric
states as those have no known efficient preparation circuits. We consider preparation circuits as
discussed in Sec.~\ref{probeStatesSec} and assume that every single-qubit gate undergoes a depolarising
noise with probability $10^{-4}$ while two-qubit gates are effected by a larger depolarisation probability
$10^{-3}$. We note that this is comparable to currently existing technology \cite{oxford_ion_trap}.
Fig.~\ref{withnoise} (solid lines) illustrates the performance of the different families of quantum
states when their preparation circuits are imperfect. Note that ansatz circuits from Sec.~\ref{probeStatesSec}
are deeper than preparation circuits of GHZ states. The performance of these ansatz states therefore
decrease more substantially due to gate noise -- compare solid to dashed lines. Nevertheless,
the superior performance of ansatz states persist even under this realistic experimental noise.


\section*{References}

\end{document}